\documentclass[sigconf,prologue,table,xcdraw]{acmart}
\renewcommand\footnotetextcopyrightpermission[1]{}  
\usepackage{hyperref}    
\usepackage{booktabs}    
\usepackage{multirow}    
\usepackage{tabularx}    
\usepackage{caption}     
\usepackage{subcaption}  
\usepackage{titlesec}    
\usepackage{enumitem}    

\newcommand{\changed}[1]{\textcolor{black}{#1}}

\titlespacing*{\section}{0pt}{1ex}{0.5ex}
\titlespacing*{\subsection}{0pt}{0.8ex}{0.4ex}
\titlespacing*{\subsubsection}{0pt}{0.6ex}{0.2ex}

\setlist{nosep}
\setlist[itemize]{topsep=0pt, partopsep=0pt, parsep=0pt, itemsep=0pt}
\setlist[enumerate]{topsep=0pt, partopsep=0pt, parsep=0pt, itemsep=0pt}

\renewcommand{\arraystretch}{0.95} 

\AtBeginDocument{%
  }

\begin{document}

\title{Quantifying Security Vulnerabilities: A Metric-Driven Security Analysis of Gaps in Current AI Standards}

\author{
Keerthana Madhavan, 
Abbas Yazdinejad, 
Fattane Zarrinkalam, 
Ali Dehghantanha
}
\affiliation{
    \institution{University of Guelph}
    \city{Guelph}
    \state{Ontario}
    \country{Canada}
}
\email{{kmadhava, ayazdine, fzarrink, adehghan}@uoguelph.ca}

\renewcommand{\shortauthors}{}

\begin{abstract}
As AI systems increasingly integrate into critical infrastructure, their security implications within AI compliance standards demand urgent attention. This paper conducts a comprehensive security audit and quantitative risk analysis of three prominent AI governance frameworks: NIST AI RMF 1.0, UK’s AI and Data Protection Risk Toolkit, and the EU’s ALTAI. We employ a novel methodology that combines a rigorous line-by-line audit process, performed by five researchers and validated by four industry experts, with a quantitative risk assessment framework. We develop metrics such as the Risk Severity Index (RSI), Attack Vector Potential Index (AVPI), Compliance-Security Gap Percentage (CSGP), and Root Cause Vulnerability Score (RCVS) to quantify security concerns in these standards. This analysis identifies 136 distinct concerns across the frameworks, revealing significant gaps between compliance and actual security. The NIST framework leaves 69.23\% of identified risks unaddressed, ALTAI demonstrates the highest vulnerability to attack vectors with an AVPI of 0.51, and the ICO AI Risk Toolkit exhibits the largest compliance-security gap, with 80.00\% of its high-risk concerns remaining unresolved. Our root cause analysis, quantified through RCVS, identifies under-defined processes (average RCVS of 0.33 for ALTAI) and insufficient implementation guidance (average RCVS of 0.25 for NIST and ICO) as major contributors to these vulnerabilities. This research offers actionable insights for policymakers and organizations implementing AI systems, emphasizing the urgent need for more robust, specific, and enforceable security controls within AI compliance frameworks. We provide targeted recommendations to enhance the security posture of each standard, bridging the gap between compliance and genuine security in AI governance. Supporting code is anonymously available at \url{https://anonymous.4open.science/r/Quantifying-AI-Standards-Risks-C45F}.
\end{abstract}



\maketitle

\section{Introduction}
Artificial Intelligence (AI) has become integral to sectors such as healthcare, finance, 
and transportation, transforming industries through enhanced operational efficiency, 
innovation, and decision-making \cite{Perifanis2023InvestigatingReview}. However, the rapid 
integration of AI into critical infrastructure introduces new security vulnerabilities, 
including model poisoning, data leakage, adversarial attacks, and malicious inputs 
\cite{LindsayPoling2024AITeams, Kaur2023ArtificialDirections}. For instance, adversarial 
attacks have led to misclassifications in autonomous vehicles, posing significant safety 
risks \cite{Boltachev2023PotentialModels}, while data poisoning has compromised healthcare 
AI models, resulting in misdiagnoses and flawed treatment recommendations 
\cite{Fernandez2024SecurityBattlefield}. The urgency of addressing these risks is underscored 
by the AI Incident Database, which reported a 55-incident increase in AI-related security 
breaches in 2023, marking a notable rise in vulnerabilities \cite{2023Study:InsideBIGDATA}.

While existing AI compliance standards—such as \textit{NIST AI RMF 1.0}, \textit{ICO AI Risk Toolkit}, and the \textit{European Commission’s ALTAI}—provide guidance on risk management, privacy, and ethics, they often fail to explicitly address security vulnerabilities \cite{Habbal2024ArtificialDirections}. 
Nevertheless, many organizations adopt these frameworks as quasi-security guides, assuming compliance ensures protection—a premise that has not been fully tested \cite{Folorunso2024CorrespondingCybersecurity, AIStudies}. In practice, the ICO AI Risk Toolkit and ALTAI were originally designed to safeguard rights or promote trustworthy AI, rather than implementing 
stringent security controls. Our analysis highlights that this broader usage \emph{introduces gaps}, since these frameworks do not systematically cover ai-specific threats—potentially leaving AI systems vulnerable \cite{Alzubaidi2023TowardsRequirements, Barletta2023AAI}. We do not suggest these frameworks fail at their original missions; rather, we expose a \emph{mismatch} between organizations’ reliance on them for security and their actual scope, 
where security is treated more as a peripheral concern. This gap exposes organizations to 
financial loss, reputational damage, and operational risks, since “compliance alone” does 
not necessarily guard against sophisticated AI-specific threats such as adversarial attacks 
and model tampering \cite{Blood2023ReliabilitySystems, Xia2023TowardsStudy}.

Existing frameworks also remain too generalized to address the nuanced security needs of AI systems, often relying on principle-based guidance that can lead to inconsistencies in implementation 
\cite{Hwang2017WhyNon-compliance, Hibbert2012SMBsCompliance}. Recent analysis show these standards lack robust countermeasures for AI-specific threats, leaving unresolved gaps such as vague definitions, 
unenforceable security controls, and insufficient direction on managing third-party AI components \cite{Bengio2024GovernmentRisks, Finocchiaro2023TheIntelligence, Taeihagh2021GovernanceIntelligence}.
Motivated by these gaps, we pose the following central research question: \textit{How effectively do current AI compliance standards protect against AI-specific threats when adopted as security guidance?} To investigate this, we conducted a \emph{line-by-line audit} of three globally recognized standards—\textit{NIST AI RMF 1.0}, \textit{ICO’s AI and Data Protection Risk Toolkit}, and \textit{ALTAI}—identifying \textbf{136 distinct security concerns}. Our analysis revealed systemic issues including ambiguous specifications, insufficient data protection measures, and challenges in enforcing security controls, particularly for third-party AI components and unforeseen uses of AI systems.

\textbf{This research makes the following key contributions:}
\begin{itemize}
    \item We present an audit framework specifically designed to identify security vulnerabilities 
          in AI compliance standards, offering a structured and repeatable process for evaluating 
          compliance documents.
    \item We introduce four new metrics—Risk Severity Index (RSI), Root Cause Vulnerability Score (RCVS), 
          Attack Vector Potential Index (AVPI), and Compliance-Security Gap Percentage (CSGP)—that provide 
          quantitative measures of security effectiveness, enabling cross-framework comparisons.
    \item We identified and categorized \textbf{136 distinct security concerns} in the NIST AI RMF, 
          ICO AI Risk Toolkit, and ALTAI standards, uncovering under-defined processes, ambiguous 
          guidance, and unenforceable controls.
\end{itemize}

These contributions provide actionable insights for auditors, policymakers, and organizations 
implementing AI systems. By quantitatively measuring the security robustness of AI compliance 
standards, this research exposes critical shortcomings in existing frameworks, underscoring 
the urgent need for more specific, prescriptive requirements to safeguard AI systems effectively. 

The remainder of this paper is structured as follows:
\textbf{Section~\ref{sec:existing_ai_security}} reviews related work on AI security and compliance.
\textbf{Section~\ref{sec:methodology}} details our audit methodology and the newly introduced metrics.
\textbf{Section~\ref{sec:audit_results}} presents the results of the audit, including a classification 
of security concerns by root cause and risk level.
\textbf{Sections~\ref{sec:ICO}, \ref{sec:altai}, and \ref{sec:nist}} provide detailed case studies 
of the ICO, ALTAI, and NIST frameworks, respectively.
\textbf{Section~\ref{sec:quantitative_analysis}} offers a comparative quantitative analysis of the 
three standards.
\textbf{Section~\ref{sec:discussion_recommendations}} discusses the implications of our findings 
and presents policy recommendations.
Finally, \textbf{Section~\ref{sec:conclusions}} concludes with a summary of our findings and outlines 
potential future research directions.
\section{Related Work}
\label{sec:existing_ai_security}
Efforts to regulate AI through standards have been extensive, yet no established mandatory standards exist. The European Commission's risk-based approach through the AI Act aims to ensure AI systems are safe, transparent, and adhere to fundamental rights \cite{Laux2022TrustworthyRisk}. Systematic studies, such as those by Xia B et al., point out that the ability of these frameworks to assess and mitigate AI risks is not well understood \cite{Xia2023TowardsStudyb}. This is further evidenced by research identifying challenges developers face in industrial fields, including ambiguous terminologies, lack of domain-specific concreteness, and non-specific requirements \cite{Hwang2022BridgingDevelopment}.

This indicates that existing standards and regulations alone may not guarantee AI system security. When organizations focus solely on compliance, potential vulnerabilities not explicitly addressed by the standards can be overlooked. This overemphasis on compliance often results in a false sense of security, leaving systems vulnerable \cite{Schneck2019FROMLaws, 2022ComplianceKey}. Moreover, simple compliance may not guarantee protection, as evidenced by a study identifying 148 issues of varying severity across three major digital compliance standards \cite{Stevens2020ComplianceStandards}. Key issues included insufficiently specific security requirements, lack of guidelines for rapid response to threats, and absence of provisions for ongoing system monitoring. These findings underscore the need for our proposed line-by-line audit to offer a more holistic understanding of security gaps and to guide the development of more robust AI compliance standards.

Additionally, selecting appropriate security standards and extracting requirements within an organizational context can be challenging. The complexity of finding the most suitable cybersecurity solutions for an organization is underscored by a study of public organizations in Ecuador \cite{Hamdani2021CybersecuritySystem}. This complexity implies that compliance alone may not ensure the security of AI systems. Consequently, organizations may need to adopt tailored security measures beyond compliance to manage their risks effectively. Historically, compliance audits have used reward-driven and penalty-based approaches to promote adherence to minimum security standards. However, these approaches may not encourage organizations to implement additional security measures beyond the basic requirements for compliance \cite{Cheng2013UnderstandingTheory, Ifinedo2012UnderstandingTheory}. This suggests that while compliance standards are needed for maintaining a baseline level of cybersecurity, they may not be sufficient to ensure complete protection. Hence, organizations need to continuously assess their specific risks, adopt relevant controls, and monitor the effectiveness of their cybersecurity programs.

Existing research indicates that although current frameworks aim to address security risks in AI systems, there is still no consensus on how to effectively define and evaluate these risks \cite{Steimers2022SourcesSystems, MartinezStudyAI}. Anderljung et al. point out that we currently lack a robust and comprehensive set of evaluation methods to operationalize these standards, which are necessary to identify and mitigate the potentially dangerous capabilities and emerging risks associated with advanced AI systems \cite{Anderljung2023FRONTIERSAFETY}. This issue is further exacerbated by the absence of concrete solutions or structured formats for presenting these evaluations \cite{Goldsteen2022AnModels, Kazim2020AFramework}.

While previous studies have examined the effectiveness of AI compliance standards, they often fall short of providing a detailed, line-by-line analysis of these standards to determine whether the existing controls are sufficient to ensure AI security. Such an analysis is crucial for uncovering potential gaps and oversights in security measures, and for evaluating whether the current standards provide adequate safeguards against emerging AI-specific threats. Our research addresses this critical gap by conducting a comprehensive, line-by-line examination of leading AI compliance standards. This approach allows us to assess the adequacy and effectiveness of existing controls, identify potential security vulnerabilities, and determine whether these standards provide sufficient guidance to ensure the security of AI systems in practice. By doing so, our study aims to highlight potential security risks that might be overlooked in more general analysis and to evaluate whether current compliance standards are truly fit for purpose in the rapidly evolving landscape of AI security.

To address the gaps identified in existing research, we developed a novel methodology that combines a detailed security audit of AI compliance standards with quantitative risk assessment. The following section outlines our approach.

\section{Research Methodology}
\label{sec:methodology}
The study, conducted between May 2023 and May 2024, received ethical approval from the University Research Ethics Board. Participants provided informed consent, and data confidentiality was maintained.

\subsection{Selection of compliance standards}
Our selection of compliance standards was based on a systematic review of 16 globally recognized AI frameworks identified by Xia et al. (2023) \cite{Xia2023TowardsStudyb}. We focused on frameworks developed between 2016 and 2023 by leading technology companies, government agencies, and industry consortia, evaluating them on risk assessment guidance, alignment with Responsible AI principles, technical depth, and regional relevance. Selection criteria included applicability to our research needs, global recognition, and distinct perspectives on security aspects. The AI compliance standards selected for our audit are NIST AI RMF 1.0 (2023) \cite{U.SDepartmentofCommerce2023Artificial1.0}, providing global guidance for managing AI system risks; UK's AI and Data Protection Risk Toolkit (2020) \cite{2023UKEnforcement}, focusing on data protection; and EU's ALTAI (2020) \cite{Stahl2022AssessingALTAI}, ensuring ethical AI use within the EU. These were chosen for their comprehensive coverage of global, national, and regional perspectives on AI compliance. For a detailed analysis of each standard, refer to Appendix \ref{app:selection_standards}.

\subsection{Participant Recruitment}

This study engaged nine participants with extensive experience in cybersecurity, compliance, and risk management across academia, industry, and government sectors. The research team comprised five researchers and four industry experts, with an average of 22.5 years of professional experience. We employed purposive sampling to recruit five researchers from diverse backgrounds \cite{Campbell2020PurposiveExamples}. These researchers conducted a detailed audit of three AI standards, following the process described in Section \ref{tab:audit_process}. To validate our findings, we recruited four Subject Matter Experts (SMEs) from industry. We outline the expert validation process and criteria in Section \ref{tab:expert_process}.

For detailed participant information and recruitment methods, refer to Appendix \ref{app:participant_details}.

\subsection{Audit Methodology for AI Compliance Standards}
We refined the systematic auditing approach by Stevens et al. (2020) to identify security issues in AI compliance standards, enhancing it with a quantitative risk assessment framework \cite{Stevens2022AboveMandates}. This approach offers advantages over traditional methods like random sampling or purely qualitative analysis, which may overlook nuances or lack structural rigor \cite{Collins2018TheResearch, Sutton2015QualitativeManagement}. By focusing on security concerns and integrating quantitative metrics, our audit addresses a gap in existing research that often examines standards primarily for trust, privacy, and ethics aspects. The methodology involves a line-by-line audit of selected AI compliance standards, followed by expert validation and quantification. This ensures both accuracy and an objective evaluation of security risks. For each identified security concern, we provide a detailed explanation, specific examples from the standards, quantitative risk assessment, and relevant real-world incidents that illustrate potential consequences. This comprehensive approach helps identify potential security vulnerabilities that might be missed in broader analysis and demonstrates their practical implications through historical precedents.

\subsection{AI Compliance-standard Audit Process}
\label{tab:audit_process}
\textbf{Audit Objective}: The primary goal of this audit was to identify potential security concerns within AI compliance standards that could undermine the security of AI systems. The audit focused on policies that present risks of data exposure and processes characterized by unclear implementation guidelines. To achieve this, a detailed line-by-line analysis of three prominent AI compliance standards was conducted. In this study, a ``security concern" is defined as any recommendation or policy that, if implemented as written, could compromise security measures, potentially leading to unauthorized access or the exposure of sensitive data.

\textbf{Audit Process}: The audit commenced with independent analysis conducted by each researcher. These analysis were guided by the audit's objective and employed a content analysis methodology grounded in established social science research principles \cite{Weber1990BasicAnalysis}. Researchers documented their findings at the conclusion of each section within the standards under review. Each documented issue included the title of the section where it was identified, the specific phrase or provision deemed problematic, a brief description of the issue, and, where applicable, references to publicly known issues. In instances where multiple issues were identified within a single phrase or section, each issue was logged separately to ensure comprehensive coverage.

Upon completing the standards examination, the researchers flagged issues based on specific criteria. An issue was flagged as (1) if it was independently identified by multiple researchers. Alternatively, an issue was flagged as (2)  if there was disagreement among the researchers regarding its significance. In cases where no unanimous consensus could be reached, the issue was discarded from the final list of concerns but maintained as a record of disagreement. This approach ensured that all perspectives were considered while maintaining the integrity of the final audit outcomes.

To validate the consistency of the findings, inter-coder reliability was calculated using Krippendorff’s $\alpha$ (Alpha), a statistical measure that accounts for chance agreement \cite{Bailey1996DesigningResearch}. This metric was chosen for its suitability in evaluating agreement in nominal data, particularly when categorizing data points based on the presence or absence of a security concern. The analysis yielded inter-coder reliability values of 0.88 for the NIST AI RMF 1.0, 0.84 for the ICO AI Risk Toolkit, and 0.90 for the ALTAI standard. An $\alpha$ value of 0.8 or higher indicates a high level of reliability in the audit process, effectively mitigating the likelihood of chance agreements among researchers and confirming the consistency of identified security concerns.

Following the verification of identified issues, the research team proceeded to analyze and categorize these security concerns through an iterative open coding process. This process involved the application of categorical labels—referred to as a ``codebook"—to the data \cite{Namey8Methods}. Each AI compliance standard was systematically coded to identify the specific security concern, its perceived root cause, the probability of its occurrence, and the severity of its potential impact. Any discrepancies among coders were resolved through collaborative discussion, resulting in the development of a stable codebook. The definitions for coded terms were established through unanimous agreement, with many terms adapted from the CRM (Composite Risk Management) framework \cite{JerryD.Vanvactor2007RiskEBSCOhost}.

The finalized \textbf{codebook} categorized root causes into four distinct types as shown in Table \ref{tab:codebook}.

\begin{table}[h!]
\centering
\tiny
\caption{Categorization of root causes in the codebook}
\label{tab:codebook}
\arrayrulecolor{gray!50}
\setlength{\tabcolsep}{3pt}
\renewcommand{\arraystretch}{1.1}
\begin{tabularx}{\columnwidth}{|p{2.5cm}|X|}
\hline
\rowcolor{gray!20} \textbf{Category} & \textbf{Description} \\
\hline
Data Vulnerability & Critical issue that could lead to data breaches or compromise the security of sensitive information. \\
\hline
Unenforceable Security Control & Control that, as written, cannot be effectively enforced and therefore requires rewording or removal from the compliance standard. \\
\hline
Under-defined Process & Absence of necessary instructions or details required for secure implementation, leading to potential security gaps. \\
\hline
Ambiguous Specification & Vague or unclear description of implementation details, which could result in varying interpretations and potentially lead to inappropriate actions or inactions. \\
\hline
\end{tabularx}
\end{table}

For the purposes of \textbf{risk analysis}, the probability and severity of each identified security concern were assessed during the audit process. The categories for probability and severity are summarized in Table \ref{tab:probability_severity_categories}. The qualitative data collected here serves as the foundation for the subsequent quantitative analysis, where we calculate key metrics that provide an objective comparison of the standards' security gaps.

\begin{table}[h!]
\centering
\tiny  
\caption{Summary of Probability and Severity Categories}
\label{tab:probability_severity_categories}
\arrayrulecolor{gray!50} 
\setlength{\tabcolsep}{3pt} 
\renewcommand{\arraystretch}{1.1} 
\begin{tabularx}{\columnwidth}{|p{2.3cm}|X|}
\hline
\rowcolor{gray!20} \multicolumn{2}{|c|}{\textbf{Probability Categories}} \\ \hline
\textbf{Category} & \textbf{Description} \\ \hline
\textbf{Frequent} & Events occurring consistently \\ \hline
\textbf{Likely} & Events expected to occur multiple times \\ \hline
\textbf{Occasional} & Events occurring sporadically \\ \hline
\textbf{Seldom} & Events that are unlikely but possible \\ \hline
\textbf{Unlikely} & Events not expected to occur \\ \hline
\rowcolor{gray!20} \multicolumn{2}{|c|}{\textbf{Severity Categories}} \\ \hline
\textbf{Category} & \textbf{Description} \\ \hline
\textbf{Catastrophic} & Complete system loss, full data breach, or comprehensive data corruption \\ \hline
\textbf{Critical} & Significant system damage or substantial data breach \\ \hline
\textbf{Moderate} & Minor system damage or partial data breach \\ \hline
\textbf{Negligible} & Minor system impairment \\ \hline
\end{tabularx}

\end{table}

Using a \textbf{risk assessment} matrix adapted from the CRM framework (refer to Table \ref{tab:risk_matrix} in Appendix \ref{app:visualizations}), the impact level of each issue was calculated as a function of both probability and severity. The resulting impact levels were classified into four tiers: extremely high, high, medium, or low. The findings from this audit, organized by the identified root causes and evaluated through the risk analysis framework, are detailed in the subsequent sections (\ref{sec:ICO}, \ref{sec:altai}, \ref{sec:nist}). 

\subsection{Risk Quantification Framework}
\label{sec:quantitative_framework}

We present a quantitative risk framework designed to provide objective and comparable measures of security across AI compliance standards. This framework quantifies vulnerability severity, identifies root causes, and evaluates the robustness of each standard, forming the foundation for our comparative analysis and recommendations for enhancing AI compliance.

Our framework is grounded in established risk management principles, including \textit{NIST SP 800-30}, \textit{ISO 31000}, and \textit{Composite Risk Management (CRM)}. Risk is quantified using a probability-impact approach, with probability values ranging from 1 (Unlikely) to 5 (Frequent) and severity values from 1 (Negligible) to 4 (Catastrophic). These four impact levels align with widely accepted risk assessment models\footnote{The impact levels of Negligible, Moderate, Significant, and Catastrophic are commonly used in risk frameworks such as CRM and ISO 31000.}, ensuring a consistent and interpretable foundation for risk quantification.

Using the qualitative audit process described in Section \ref{tab:audit_process}, we compute key risk metrics based on these probability and severity values. Central to this framework is the Risk Score (RS) assigned to each identified concern\footnote{In this context, 'concerns' refer to the individual items listed in the 'concern' column of the 'ai\_compliance\_auditv2\_pyf' Excel sheet, available in the associated repository.}. The RS supports the calculation of quantitative metrics that drive our analysis.

\subsubsection{Risk Score (RS)}: 
\changed{Each concern $i$ is assigned a Risk Score (RS) as:
\begin{equation}
   RS_i = Probability_i \times Impact_i
\end{equation}
where $Probability_i \in \{1,2,3,4,5\}$ represents the likelihood of occurrence and $Impact_i \in \{1,2,3,4\}$ represents the severity of the concern.}

\subsubsection{Risk Severity Index (RSI)}: 
\changed{The Risk Severity Index (RSI) quantifies overall risk exposure across all identified concerns:
\begin{equation}
   RSI = \frac{\sum_{i=1}^{n} RS_i}{n}
\end{equation}
where $n$ represents the total number of concerns. RSI provides a concise measure of the overall risk severity in a framework, supporting cross-framework comparisons.
}
\subsubsection{Root Cause Vulnerability Score (RCVS)}:  
\changed{The Root Cause Vulnerability Score (RCVS) quantifies the contribution of each root cause to the overall risk:
\begin{equation}
   RCVS_c = \frac{\sum_{i \in C_c} RS_i}{\sum_{i=1}^{n} RS_i}
\end{equation}
where:
\begin{itemize}
   \item $\sum_{i \in C_c} RS_i$ is the sum of risk scores for all concerns in category $c$.
   \item $\sum_{i=1}^{n} RS_i$ is the total risk score for all concerns.
\end{itemize}
}

This metric identifies the root causes (e.g., under-defined processes, ambiguous specifications) that contribute most significantly to overall risk\footnote{Root causes refer to categories of vulnerabilities, such as under-defined processes, ambiguous specifications, and data vulnerabilities. These categories are derived during the qualitative analysis phase.}. Categories with higher Root Cause Vulnerability Scores (RCVS) indicate a greater impact on total risk.

\subsubsection{Attack Vector Potential Index (AVPI)}:  
\changed{The Attack Vector Potential Index (AVPI) measures how unresolved vulnerabilities from root causes contribute to the system's attack surface:
\begin{equation}
   AVPI = \sum_{c=1}^{k} \left( \frac{|C_c|}{|C_{\text{total}}|} \cdot RCVS_c \right)
\end{equation}
where:
\begin{itemize}
   \item $|C_c|$ is the number of concerns in root cause category $c$.
   \item $|C_{\text{total}}|$ is the total number of concerns.
   \item $RCVS_c$ is the Root Cause Vulnerability Score for category $c$.
   \item $k$ is the number of distinct root cause categories.
\end{itemize}
}
The Attack Vector Potential Index (AVPI) measures the system's exposure to potential attack vectors\footnote{An attack vector is the path or method used by an attacker to exploit a vulnerability. A higher AVPI indicates greater exposure and a higher likelihood of exploitation.

A higher AVPI reflects a larger attack surface, as unresolved vulnerabilities from specific root causes increase system exposure.}

\subsubsection{Compliance-Security Gap Percentage (CSGP)}:  
\changed{The Compliance-Security Gap Percentage (CSGP) quantifies the share of high-risk and extremely high-risk concerns that remain unaddressed:}
\begin{equation}
   CSGP = \frac{|C_{\text{unaddressed}}|}{|C_{\text{total}}|} \times 100
\end{equation}
where:
\begin{itemize}
   \item $|C_{\text{unaddressed}}|$ is the number of concerns classified as High (H) or Extremely High (E) risk.
   \item $|C_{\text{total}}|$ is the total number of concerns.
\end{itemize}

Concerns are classified as Extremely High (E) when:
\begin{equation*}
\begin{split}
   & (\text{Probability} \in \{4,5\} \land \text{Severity} = 4)~\lor \\
   & (\text{Probability} = 5 \land \text{Severity} \in \{3,4\})
\end{split}
\end{equation*}

Concerns are classified as High (H) when:
\begin{equation*}
\begin{split}
   & (\text{Probability} \in \{3,4\} \land \text{Severity} = 3)~\lor \\
   & (\text{Probability} \in \{2,3\} \land \text{Severity} = 4)~\lor \\
   & (\text{Probability} \in \{4,5\} \land \text{Severity} = 2)
\end{split}
\end{equation*}

\changed{A higher CSGP indicates a larger share of unresolved high-risk concerns, reflecting gaps in the compliance framework.}

\subsubsection{Justification of the Quantification Framework}
The quantification framework is based on principles from \textit{NIST SP 800-30}, \textit{ISO 31000}, and \textit{Composite Risk Management (CRM)}. Each metric addresses a critical dimension of risk assessment: the Risk Severity Index (RSI) captures overall risk severity, the Root Cause Vulnerability Score (RCVS) identifies the most impactful root causes, the Attack Vector Potential Index (AVPI) highlights systemic exposure to attacks, and the Critical Severity Gap Percentage (CSGP) quantifies unresolved high-risk concerns. By employing a probability-impact approach with well-defined risk parameters, the framework ensures computational efficiency, scalability, and interpretability, enabling policymakers and security practitioners to effectively prioritize risk mitigation efforts.

\subsection{Expert validation process}
\label{tab:expert_process}
To obtain external validation of our findings, our \textit{four experts}, as shown in Table \ref{tab:detailed_demo}, from real-world organizations helped validate the findings from the researchers. We asked these experts to categorize the security concern we identified into one of three categories: (1) confirmed, (2) plausible, or (3) rejected. A confirmed security concern is one that the expert has previously encountered or observed its consequences within an enterprise environment. A plausible issue is one that the expert hasn't personally encountered but agrees could potentially arise in other organizations or if the controls were implemented as stated. A rejected issue is one where there's no observable evidence of security concerns in a live environment or there are related security factors we hadn't considered.

We employed both closed and open-ended survey questions to collect insights from each expert. Besides simply confirming or dismissing each identified issue, we also encouraged experts to share relevant personal experiences, adding depth to their responses. We presented the issues to the experts in a randomized order through an Excel workbook, providing the referenced section title, exact text from the section, a brief detail of the security concern, a description of the perceived issue, and the standard document. In our study, the expert validation process was governed by a consensus threshold of 75\%. For a finding to be accepted in our panel of four experts, it requires the concurrence of at least three experts, which represents 75\% agreement. Conversely, a finding where only two experts agree, representing a 50\% agreement, is rejected. This strong consensus requirement aligns with established inter-coder reliability practices, enhancing the credibility of our results \cite{Engagedscholarshipcsu2002ContentReliability}. After gathering data from each expert, we removed the rejected finding. We also held open-ended discussions with the experts to discuss similarities and differences in assessments. 

\paragraph{Expert partner criteria} We established the following criteria for partnering with organizations: (1) They are actively employed by an organization that uses digital security compliance programs. (2) Their current work role and experience with cybersecurity and compliance programs. After several negotiations, we established memorandums of understanding with the four partnering organizations that met our criteria. Leaders within each organization nominated several compliance experts; we sent each participant an email outlining the voluntary nature of the study, as well as our motivation and goals. Table \ref{tab:detailed_demo} shows the qualifications of our four voluntary experts. Experts consented (as shown in \textbf{Appendix \ref{app:c_form}}) and completed their surveys during regularly scheduled work hours and received no additional monetary incentives for participating. 

\paragraph{Security concern selection}
Our commitment to minimize disruption to the participating expert's daily responsibilities was only feasible to validate a subset of our identified security concerns. Research suggests that the quality of survey responses decreases over time, and excessive time away from work may result in an expert terminating their participation in the study \cite{Hugick2008EncyclopediaMethods}. To this end, we designed our surveys to be finished by experts within 60-110 minutes. The average time taken was approximately 64.8 minutes. Given our limited pool of experts, we had to validate only a subset of our findings that was selected semi-randomly, prioritizing extremely high-risk and high-risk concerns. While we recognize the value of full validation for all identified concerns, we believe this targeted approach, selection criteria, and efficient survey design will provide insights into our survey's most critical security areas. 

\paragraph{Pilot} Prior to deploying our survey, we conducted a pilot with three security practitioners to test the survey's relevance and clarity. Feedback from this pilot was incorporated into the finalized questionnaire available in \textbf{Appendix \ref{app:surveyquestions}}, enhancing the study's overall validity and effectiveness.

\subsection{Limitations}
Our study has several limitations that warrant consideration. Firstly, our analysis focused on three AI compliance frameworks, potentially limiting the global applicability of our findings. This approach may not generalize well to regions with different regulatory environments, cultural attitudes towards AI, or technological infrastructures. 
The expert validation process, while rigorous, involved a relatively small sample of four industry experts, which may not fully capture the complexities of real-world security issues across diverse contexts. Our methodology did not account for false negatives, possibly overlooking some security concerns. Additionally, we conducted audits in isolation, assuming flawless implementation of each standard, which may not reflect real-world scenarios where multiple security controls interact.
The subjective nature of our risk categorization, relying heavily on expert judgment, introduces potential bias. Interpretations of risk severity may vary across different contexts, highlighting the need for more standardized assessment guidelines. Lastly, the rapidly evolving nature of AI technology means that some of our findings may become outdated as new challenges emerge.
Despite these limitations, our methodology offers a robust framework for evaluating AI standards. Future research should address these constraints by expanding the scope to diverse geographic regions and industries, involving a larger expert panel, and conducting comparative studies across a broader range of AI governance frameworks.

\section{Audit Results}
\label{sec:audit_results}

\begin{figure}[ht]
  \centering
  \includegraphics[width=0.3\textwidth]{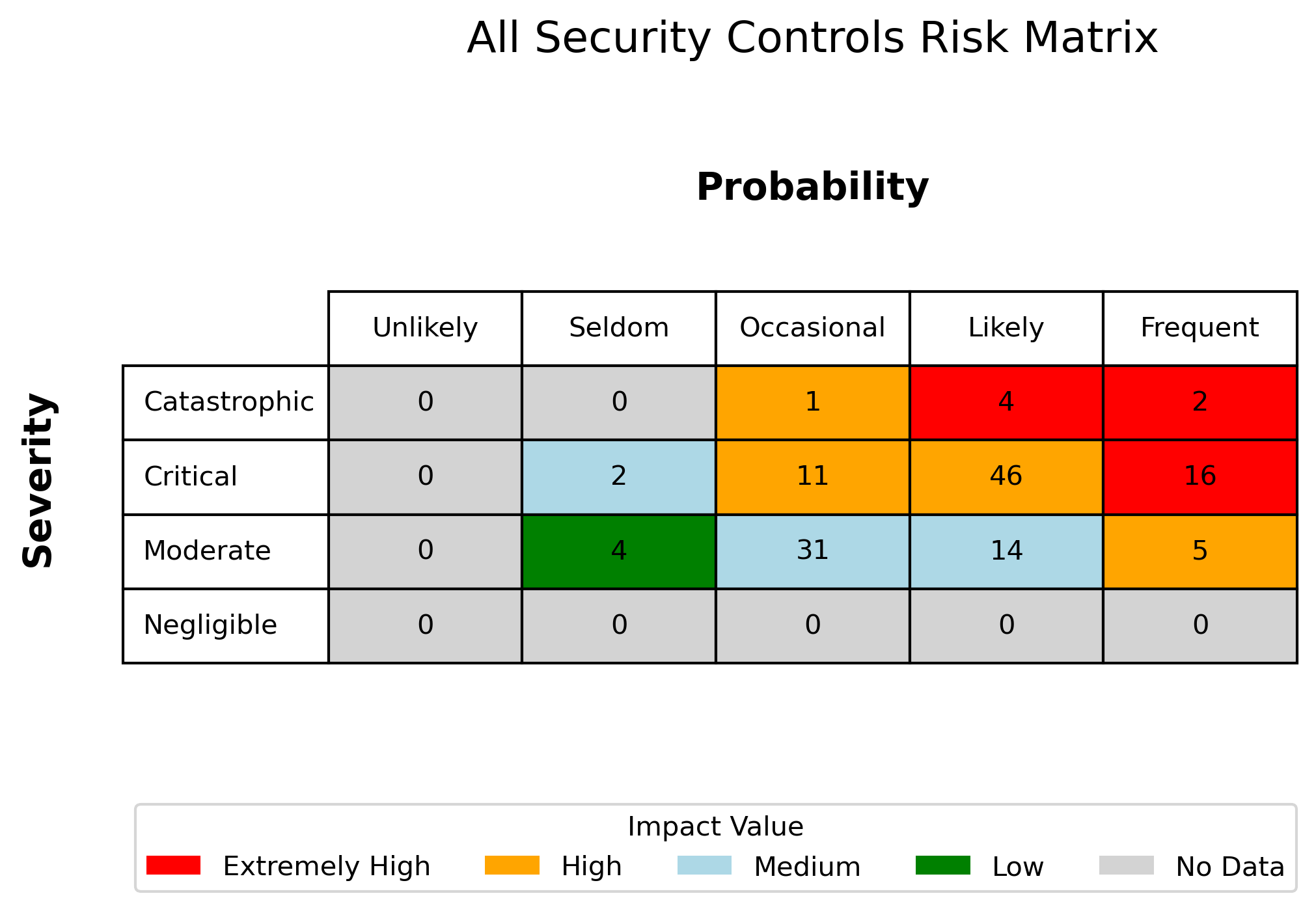}
  \caption{Risk matrix showing the impact level of all the security concerns identified across the standards}
  \label{fig:overall_matric}
\end{figure}

The audit of AI compliance standards included a thorough evaluation of three primary documents: NIST AI RMF 1.0, ALTAI HLEG EC, and the ICO AI Risk Toolkit. Across these standards, we identified a total of \emph{136} security concerns, classified by their severity and root causes. Table \ref{tab:security_concerns} provides a breakdown of these concerns by document and assessed risk levels. Following the CRM framework, these concerns are classified and assessed within a risk matrix Figure \ref{fig:overall_matric}. 

\begin{table}[h!]
\centering
\caption{Security Concerns by Document and Assessed Risk}
\label{tab:security_concerns}
\resizebox{\columnwidth}{!}{%
\begin{tabular}{|p{5cm}|c|c|c|c|c|}
\hline
\rowcolor{gray!30} \textbf{Document} & \textbf{Total Concerns} & \textbf{Extremely High} & \textbf{High} & \textbf{Medium} & \textbf{Low} \\ \hline
AI and Data Protection Risk Toolkit & 30 & 3 & 16 & 11 & 0 \\ \hline
ALTAI & 78 & 19 & 30 & 26 & 3 \\ \hline
NIST AI Risk Management & 28 & 0 & 17 & 10 & 1 \\ \hline
\end{tabular}%
}
\end{table}

The overview presented above highlights the varying levels of security concerns across the three standards. To gain a deeper understanding of these issues and their implications, we will now proceed with a detailed evaluation of each standard. These evaluations will explore the specific vulnerabilities, their root causes, and real-world examples that illustrate the potential consequences of these security gaps.

\section{Evaluation: AI and Data Protection Risk Toolkit}
\label{sec:ICO}
We identified \emph{30} security concerns within the standard text. These concerns were evaluated and classified according to their impact levels: 3 extremely high concerns, 16 high-risk concerns, and 11 medium-risk concerns. The impact level assessment was crucial in identifying and categorizing the perceived security concerns in AI systems, highlighting the potential risks and vulnerabilities in data handling and protection. Furthermore, our analysis chose to omit two incidents of unenforceable security control and ambiguous specification. Expert validation determined that these incidents did not create insecure conditions or promote insecure practices.

\begin{figure*}[htbp]
    \centering
    \begin{subfigure}[b]{0.32\textwidth}
        \centering
        \includegraphics[width=\linewidth]{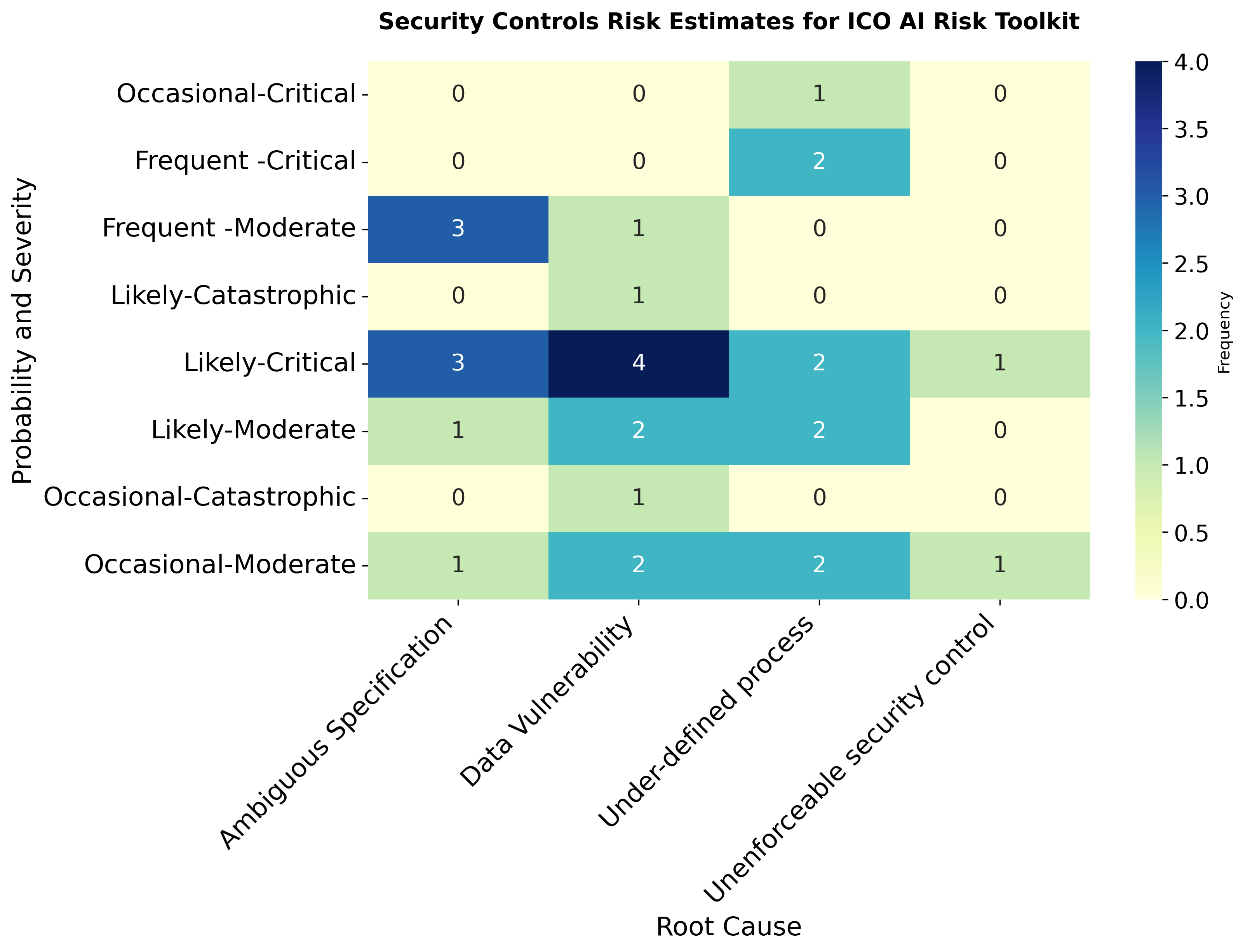}
        \caption{Correlation between root causes and risk impact.}
        \label{fig:heatmaprc}
    \end{subfigure}
    \hfill
    \begin{subfigure}[b]{0.32\textwidth}
        \centering
        \includegraphics[width=\linewidth]{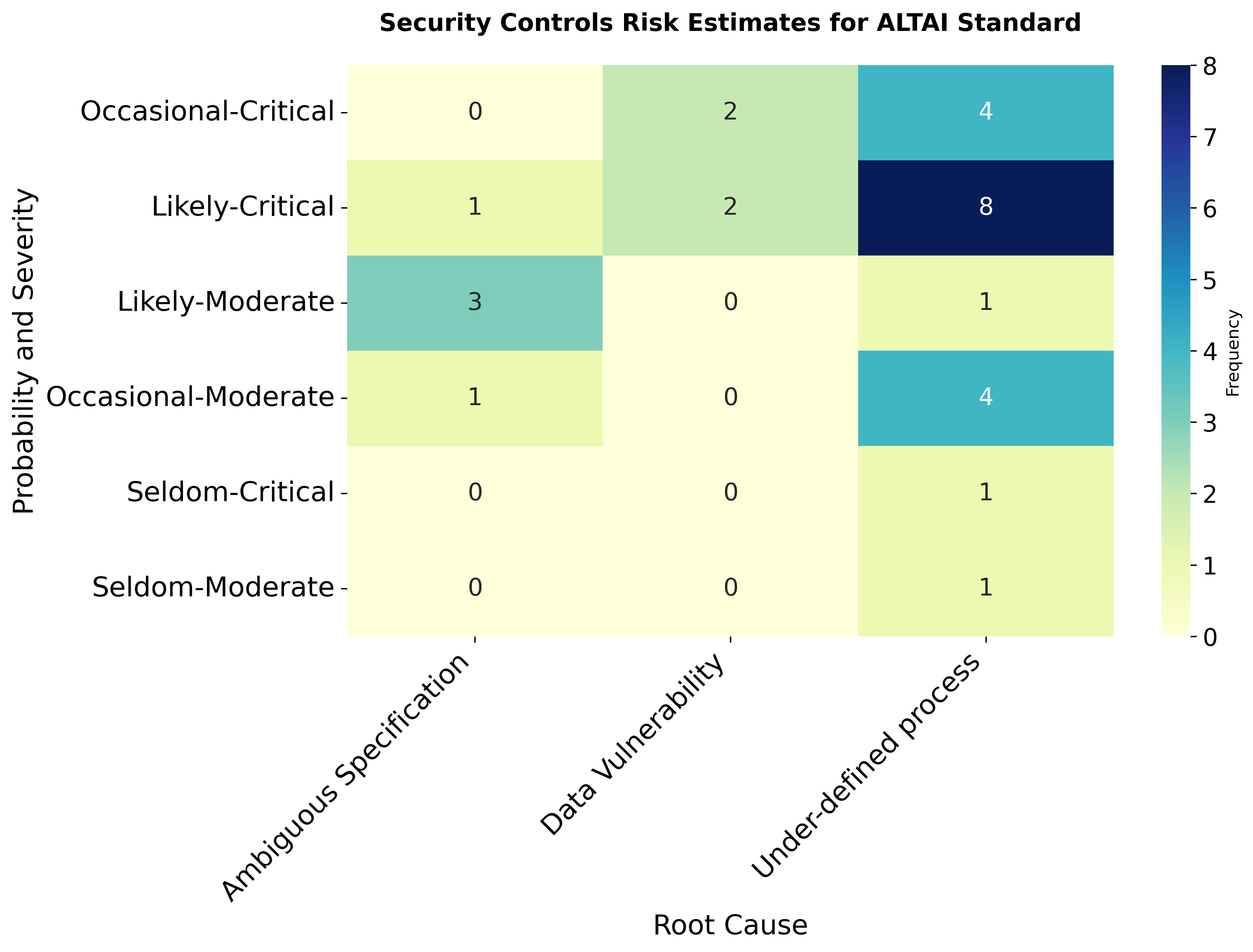}
        \caption{Correlation between root causes and risk impact (ALTAI).}
        \label{fig:figure4altai}
    \end{subfigure}
    \hfill
    \begin{subfigure}[b]{0.32\textwidth}
        \centering
        \includegraphics[width=\linewidth]{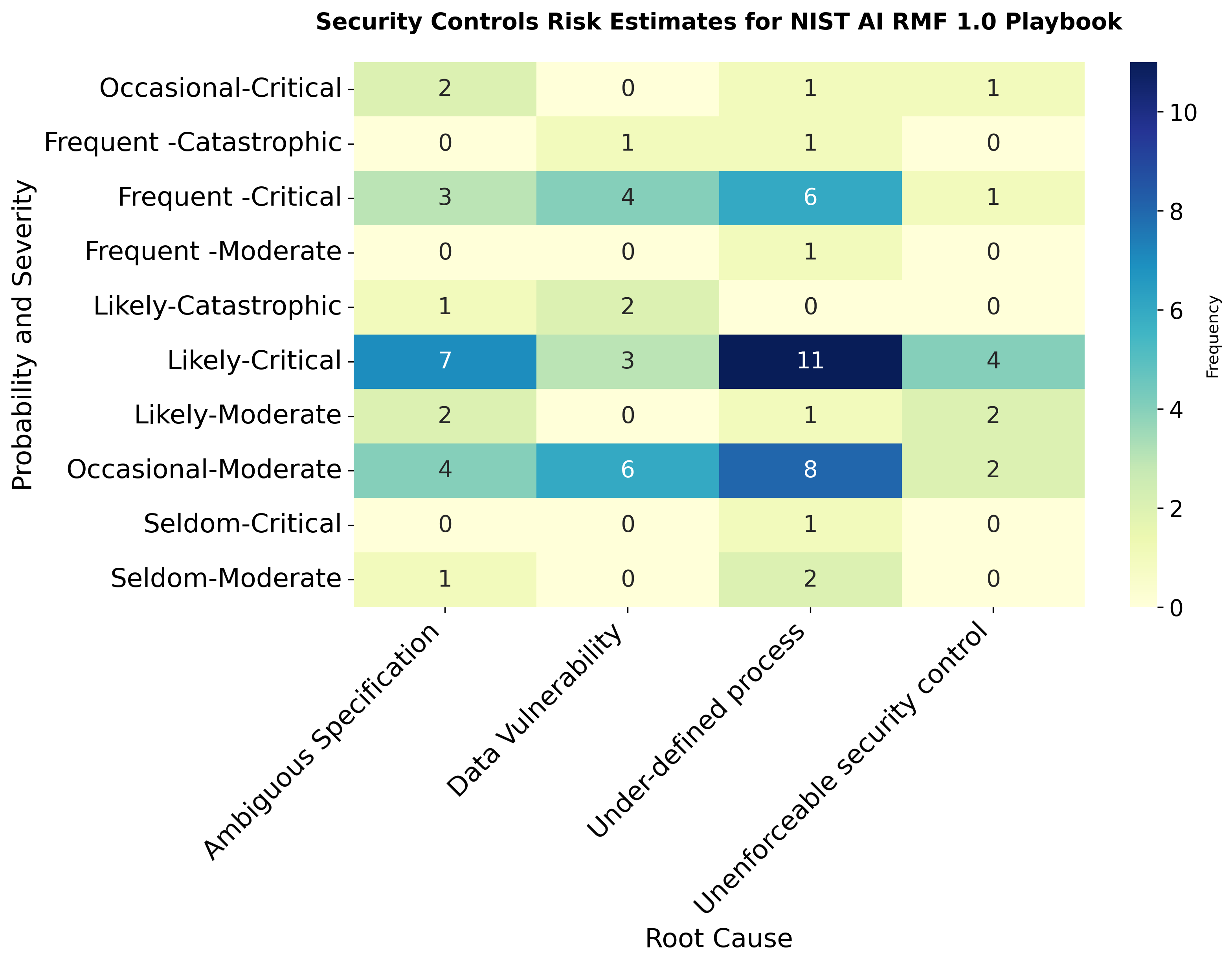}
        \caption{Correlation between root causes and their impact (NIST).}
        \label{fig:figure5}
    \end{subfigure}
    \caption{Heatmaps illustrating the correlation between root causes and risk impact across different standards.}
    \label{fig:heatmaps_combined}
\end{figure*}

Refer to Figure \textbf{\ref{fig:heatmaprc}} that further illustrates these findings. The heatmap's gradient indicates the frequency of security concerns across different impact and probability levels, with darker shades signifying more frequent occurrences. These shades highlight areas with higher event frequencies but don't necessarily correspond to higher risk levels, indicating that frequent issues can span from low to high severity. Below, we present detailed examples of findings based on their perceived root cause.

\subsection{Root cause analysis}

\subsubsection{\textbf{Data vulnerability}} The \emph{11} security concerns identified primarily stemmed from weak data flow mapping and protection measures. Section 1.3 of the standard suggests but does not mandate data flow mapping, creating a gap in data security protocols. This omission exposes systems to increased risks of data breaches, regulatory non-compliance, and privacy violations, highlighting the necessity for mandatory data mapping to ensure comprehensive data protection and adherence to regulatory standards.

\textit{Real-World examples illustrating the risks}: The security implications of inadequate data flow mapping are vividly illustrated by specific AI vulnerabilities, such as the arbitrary file write vulnerability identified in MLFlow, known as CVE-2023-6975 \cite{KevinTownsend2024EightSecurityWeek}. With a Common Vulnerability Scoring System (CVSS) severity rating of 9.8, this vulnerability underscores the dire consequences of unauthorized access and malicious activities potentially due to gaps in data flow oversight. Furthermore, the risk of poisoned training data, particularly in large language models, exemplifies how compromised data integrity can lead to biased outputs, security breaches, or complete system failures. These occurrences emphasize the importance of accurate data flow mapping in AI systems to uphold data integrity, security, and overall dependability. Addressing the complex data flows in AI requires scalable and sophisticated approaches. Tools and methodologies designed for automated data discovery and mapping, such as AIMap from Adeptia, offer viable solutions \cite{Adeptia2023AIMap}. These allow for the efficient identification and protection of data pathways. While we understand that it is not practical for an organization to map every single data flow in the system, adopting a risk-based approach enables the prioritization of essential data flows. This ensures that robust security measures are implemented where they are most needed.

\subsubsection{\textbf{Under-defined process}}
Our analysis identified \emph{9} security concerns related to the absence of clear guidelines for secure implementation. Section 1.7 of the standard merely recommends regular discussions on personal data collection without providing specific data requirements, emphasizing data minimization principles, or outlining the necessary justifications for data collection. These are elements recognized in the General Data Protection Regulation (GDPR) and other privacy frameworks to prevent unauthorized data access and ensure compliance with evolving regulatory landscapes.
The absence of precise and comprehensive processes for data collection and utilization poses privacy risks and directly contributes to security vulnerabilities. In the context of AI systems, which often process vast amounts of sensitive and personal information, this oversight can lead to inadequate protection against unauthorized access and potential data breaches. The implications of mishandling such data are profound, affecting not only privacy violations but also AI applications' ethical and societal impacts. Diligent internal oversight and discussions are emphasized to prevent data misuse and ensure responsible AI management.

\textit{Real-World Examples Illustrating the Risks}: 
In 2022, several U.S. tax filing websites, including H\&R Block, TaxAct, and TaxSlayer, used the Meta Pixel to collect and transmit sensitive financial information of taxpayers to Meta \cite{Fyler2023WhyTechHQ}. This data included names, email addresses, income, filing status, refund amounts, and more. The Meta Pixel, a piece of JavaScript code embedded in the websites, tracked user interactions and sent detailed logs to Meta, where the data was used to train AI algorithms for targeted advertising \cite{Hongsdusit2022TaxMarkup, Jennings2022ThisBoulevard}. The incident highlighted privacy concerns and the risks associated with under-defined data collection and sharing processes. The lack of clear internal guidelines and oversight allowed the Meta Pixel to collect and transmit sensitive information without proper user consent, leading to potential violations of privacy laws and regulations. 

\subsubsection{\textbf{Ambiguous specification}}
 We have pinpointed \emph{8} security concerns rooted in the vagueness of the control measures prescribed by the standard, especially evident in sections that overlook the practical application of data collection and processing guidelines. For example, Section 2.2's vague recommendation for ``appropriate" technical measures for bias mitigation lacks the precision needed for effective implementation. Similarly, the requirement in Section 2.7 for information to be ``easily accessible and easy to understand" introduces subjectivity, lacking a consistent benchmark for ease and accessibility. This ambiguity in specifications can result in a wide range of interpretations, leading to uneven application and the potential for non-compliance with regulatory standards. Without clear guidelines, stakeholders are forced to interpret the requirements, navigating the gray areas of data handling and processing. This situation escalates the risks of bias, privacy violations, and ensuing legal challenges.

\textit{Real-World Examples Illustrating the Risks}: The controversy around Microsoft's AI chatbot Tay in 2016 is a direct consequence of ambiguous data collection parameters, culminating in the bot generating offensive content after interacting with a particular user subpopulation \cite{2016TayGuardian}. This oversight, a direct consequence of ambiguous data handling protocols, allowed the bot to generate and disseminate offensive content after being exposed to harmful interactions with a subset of users. This incident underscores the need for explicit data management instructions to prevent similar misuse of technology.
Similarly, the Google+ incident 2018, where a software flaw exposed users' private data, underscores the consequences of unclear data usage policies. The flaw resulted from poorly defined security parameters within the platform's data management systems, highlighting the dire need for precise specifications in handling and protecting user data \cite{2018GoogleMashable}.

\subsubsection{\textbf{Unenforceable security control}}
We have identified \emph{2} high-risk concerns where security controls are effectively unenforceable due to vague execution details. The standard’s recommendation in Section 2.1 to consult with a data protection officer on lawful data processing bases is non-specific, potentially leading to inconsistent interpretations and applications. Section 4.6 similarly falls short by vaguely advising regular reviews of processing and privacy notices without concrete steps, risking deviation from original data purposes. 
When rules are not clear or enforceable, organizations might process data without proper authorization or fail to keep the necessary records. This can weaken efforts to protect data and comply with privacy laws. From what we've seen, relying on vague guidelines can accidentally lead organizations to break these laws or privacy norms. The original standard advice was to regularly check with a Data Protection Officer (DPO) to ensure data processing is legal. Initially, we deemed this guidance too ambiguous to be actionable, as it lacked specific implementation instructions. However, following further expert consultations, we recognize that this control can be enforceable through proper documentation of the consultation process.

\textit{Real-World Examples Illustrating the Risks:} An example of this can be seen in instances where organizations, due to unclear guidelines, fail to consult appropriately with their DPOs, leading to unauthorized data processing and breaches of privacy laws. Without clear documentation and enforceable steps, such consultations are prone to inconsistency and inadequate compliance, exposing the organization to legal and reputational risks.

\subsection{Expert recommendations}
In our expert validation process, our four recruited professionals carefully examined the findings and provided valuable feedback, resulting in additional recommendations. Each expert brought forth their perspective and insights:

\textbf{E1}, specializing in data governance, emphasizes the role of data flow mapping in identifying and mitigating potential security vulnerabilities. To transition from best practice to mandatory standard, E1 proposes a structured approach to implementing a data governance framework, leveraging advanced data discovery and classification technologies. These would facilitate the creation of precise data flow diagrams, thus providing a clear visualization of how data transits through various systems and pinpointing potential weak points that could be exploited, similar to the vulnerabilities presented in the CVE-2023-6975 incident.
\textbf{E2} brings forth an integrated strategy for data protection, combining technological solutions with policy-driven approaches. Reflecting on the gaps that led to the Meta Pixel incident, E2 recommends the establishment of a holistic Data Loss Prevention (DLP) ecosystem backed by robust encryption protocols, such as AES-256 for data at rest and TLS 1.3 for data in motion. Beyond technology, E2 stresses the importance of rigorously defined access control policies to ensure that data indices are only accessible to vetted personnel, effectively minimizing the risk of unauthorized data exposure.

With a specialization in AI security, \textbf{E3} takes a forward-looking stance, advocating for adopting AI-powered security mechanisms. These include machine learning algorithms within security information and event management (SIEM) systems that continuously learn and adapt to new threats, providing a proactive defense mechanism. E3 further advises on the strategic use of AI for data classification, which would dynamically assign sensitivity levels to data sets and determine appropriate handling procedures, aiming to curtail the kind of biases and misclassification risks exemplified by the AI anomalies seen in past large language models.
Lastly, \textbf{E4}, with an understanding of risk management, underscores the necessity for a tailored security approach. To prevent incidents akin to the Google+ data exposure, E4 proposes an alignment with established frameworks like NIST SP 800-30 for conducting comprehensive risk assessments. Crucially, E4 insists on detailed documentation practices that record consultations with Data Protection Officers (DPOs), thereby making the control measures enforceable and demonstrably integrated into the organization's operational fabric.

\section{Evaluation: ALTAI}
\label{sec:altai}
Our audit identified \emph{28} security concerns within the standard text. These concerns were evaluated and classified according to their impact levels: 17 at high risk, 10 at medium risk, and 1 at low risk. In alignment with our expert validation process, we excluded one concern about unenforceable security control. Our experts determined that machine learning experts can implement this particular control without introducing any insecure practices or compromising the overall security measures. Refer to Figure \textbf{\ref{fig:figure4altai}} for a visual representation of the correlation between the root cause and its respective probability and severity. 

Below, we present the discovered ALTAI Security concerns, categorized by root cause, and provide detailed explanations. 

\subsection{Root cause analysis}
\subsubsection{\textbf{Under-defined processes}}
We have identified \emph{19} security concerns originating from under-defined processes in AI systems. These issues stem from a combination of factors: a lack of transparency, leading to user confusion; an excessive dependence on AI in fields like healthcare, where its decision-making can be obscure and problematic; an absence of adequate oversight and control, resulting in unchecked AI operations; and poorly defined methods for evaluating AI's outputs, which poses risks due to unverified decisions.

\textit{Real-World Examples Illustrating the Risks}:
The repercussions of these under-defined processes are vividly highlighted through two cases: the UK Algorithmic Grade Prediction scandal of 2020 and the criticisms faced by IBM's Watson for Oncology in 2018 \cite{2020UKVerge, 2018IBMsOnline}.

The UK Algorithmic Grade Prediction controversy showcases the pitfalls of insufficient transparency and oversight in AI decision-making. Employing an algorithm to predict student grades without clear guidelines led to public outrage, as outcomes were perceived as unfair and biased. This incident underscores the need for transparency in AI systems, particularly when their decisions impact individuals' lives and futures \cite{Collins2021ArtificialAgenda}. Similarly, the debate around IBM's Watson for Oncology in 2018 elucidates the risks tied to an over-reliance on AI in making healthcare decisions without proper human oversight. The system's sometimes contradictory recommendations to established medical practices emphasize the dangers of operating AI systems without transparent decision-making processes and expert validation\cite{Bernal2022TransparencyWorldwide}.

These examples underscore the need for clear operational guidelines and rigorous oversight in AI systems, particularly in sensitive fields like education and healthcare. Well-defined processes are crucial for evaluating AI outputs, ensuring transparent decision-making, and deploying trustworthy systems, ultimately reducing operational risks and security vulnerabilities.

\subsubsection{\textbf{Ambiguous specification}}
Ambiguous specifications in AI systems have led to the identification of \emph{5} security concerns, particularly highlighted by the unclear boundaries around human-AI interaction and the simulation of social interactions. This ambiguity blurs the line for users between interacting with humans or AI systems and obscures the autonomous nature of AI decisions, risking the incorporation of unintended biases into the decision-making process. When AI systems lack clear specifications, users might harbor unrealistic expectations or mistrust towards these systems due to a lack of communication about their technical limitations and potential risks. Additionally, without a consistently applied definition of fairness, discrimination issues could be amplified throughout the AI lifecycle.

\textit{Real-World Examples Illustrating the Risks}:  The repercussions of these ambiguities are starkly evident in cases like Amazon's recruiting tool and Optum's healthcare algorithm. Amazon's tool, which utilized unclear criteria for evaluating candidates, perpetuated gender biases by systematically undervaluing female applicants \cite{Iriondo2018AmazonUniversity, Dastin2018InsightReuters}. This case highlights how ambiguity in AI specifications can lead to direct, systematic discrimination. Similarly, the healthcare algorithm developed by Optum, by not clearly defining healthcare needs, inadvertently skewed resource allocation in favor of certain racial groups over others \cite{ObermeyerDissectingPopulations, 2019WidelyHealth}. This serves as an example of how vague AI specifications can result in unfair resource distribution and reinforce societal biases.

\subsubsection{\textbf{Data vulnerability}} 
We identified \emph{four} vulnerabilities related to inadequate risk assessment and security guidelines for AI systems, as highlighted in Technical Robustness and Safety (Requirement 2) and Privacy and Data Governance (Requirement 3). These standards advocate for ``state-of-the-art" privacy and data protection yet lack clear definitions and implementation guidance. This vagueness can lead to superficial risk assessments and inadequate security measures, failing to address AI-specific vulnerabilities.

\textit{Real-world Examples Illustrating the Risks: }
The consequences of these inadequacies are not merely theoretical but have manifested in real-world exploits that highlight the urgent need for more rigorous standards. A notable example involves the exploitation of AI systems equipped with web-based APIs. Malicious actors have devised applications interacting with these APIs, launching sophisticated attacks that exploit the systems' data vulnerabilities. One such attack vector is the manipulation of inputs in a way that bypasses AI-driven image content filters. These manipulated inputs might appear benign to human observers but are designed to deceive the AI, compromising its integrity and functionality. This was starkly demonstrated in research by Comiter (2019), where AI systems were fooled by inputs crafted to exploit their inability to discern malicious alterations designed to look normal \cite{Comiter2019AttackingIt}.

\subsection{Expert recommendations}
Our experts, who have applied the ALTAI framework to their own AI systems and security measures, provided valuable insights into the practical implications of our findings. The validation process confirmed the majority of our identified security concerns. However, the experts also provided additional context and nuance, highlighting the complexity of implementing robust security measures in AI systems.

\textbf{E1} emphasizes the need for clear process definitions in AI systems, which should include detailed guidelines on user interactions, data handling, and responses to various scenarios. Regular training for stakeholders is important to ensure they fully understand these processes, addressing problems like lack of transparency and inadequate oversight that can lead to security vulnerabilities. E1 notes that companies such as Microsoft have implemented their own responsible AI governance frameworks, focusing on human-AI interaction guidelines and the importance of training stakeholders in AI operations.

\textbf{E2} and \textbf{E3} highlight the need for robust data protection measures. They advocate for implementing strong encryption, secure data storage, and processing methods alongside a lifecycle approach to data protection. This recommendation is particularly pertinent in light of vulnerabilities related to inadequate risk assessment and security guidelines. Industries such as healthcare and finance, regulated by standards like HIPAA and SR-11-7, respectively, exemplify the adoption of these measures. They employ advanced encryption and secure data processing techniques, showcasing a commitment to protecting data throughout its lifecycle, from creation to disposal.
\textbf{E4} recommends the development of enforceable security controls, including clear policies, robust access control measures, and regular monitoring systems to detect policy violations. The suggestion to use automated enforcement tools, such as policy enforcement points (PEP), ensures consistent application of security controls. This is mirrored in regulatory frameworks like Canada's Directive on Automated Decision-Making and the European Union's AI Act, which mandate regular monitoring and strict compliance with security policies for AI systems.

\section{Evaluation: NIST Artificial Intelligence Risk Management}
\label{sec:nist}

We identified a total of \emph{78} security concerns, each categorized according to root causes, impact levels, and their respective frequencies. The impact levels were classified as 19 at extremely high risk, 30 at high risk, 26 at medium risk, and 3 at low risk. In our validation process, we excluded 9 security concerns related to unenforceable security control, ambiguous specification, and data vulnerability. Our experts determined that such controls while posing potential risks, can be addressed adequately by AI and cybersecurity experts without introducing insecure practices or compromising the overall security measures.

Refer to Figure \textbf{\ref{fig:figure5}} below for a visual representation of the correlation between the root cause and its probability and severity. Below, we present the vulnerabilities discovered in NIST AI RMF, categorized by root cause.

\subsection{Root cause analysis}
\subsubsection{\textbf{Under-defined process}}

In analyzing AI system operations, we've pinpointed \emph{32} security concerns primarily stemming from under-defined processes, marking it as the predominant issue. These concerns highlight a gap in clear protocols and procedures, elevating the risk of incidents and span issues like inadequate supervision, vague quality criteria, poor logging practices, and a lack of explicit channels for third-party reporting.
The consequences of such under-defined processes are multifaceted. With third-party involvement, for instance, the failure to effectively communicate and implement policies can introduce additional vulnerabilities. This is compounded by the often unclear decommissioning processes for third-party systems, components, and models, presenting substantial risks. The lack of specificity in defining when a system or component ``exceeds risk tolerances" fosters inconsistent practices, potentially resulting in the operation of high-risk entities beyond their safe tenure. This situation poses a threat, especially when it involves exposing sensitive data, whether belonging to third parties or intrinsic to the models.

\textit{Real-World Examples Illustrating the Risks}:
Controversies around Facebook’s News Feed Algorithm and the Uber self-driving car accident illustrate the severe consequences of under-defined AI processes \cite{2022FacebookNewsEngadget, 2020UbersNews}. Facebook's algorithm, without clear guidelines, inadvertently promoted divisive and misleading content, fostering misinformation and societal division and negatively impacting public discourse. Similarly, the Uber accident, caused by unclear processes and insufficient oversight, resulted in the first pedestrian fatality involving a self-driving car. This incident underscored the dangers of automation complacency, led to the suspension of Uber’s self-driving tests, and triggered a comprehensive reassessment of safety protocols. These examples highlight the urgent need for explicit, well-defined processes in AI operations to prevent risks ranging from misinformation to life-threatening situations.

\subsubsection{\textbf{Ambiguous specification}}
Our audit identified \emph{20} security concerns related to ambiguities in terminologies or specifications, which can lead to diverse interpretations and discrepancies in security implementation and evaluation. Such ambiguities, exemplified by the undefined term ``acceptable limits", can result in varying, potentially insecure practices. Without a clear definition, different AI developers or organizations might interpret this term in vastly different ways, potentially leading to varying and potentially insecure practices. 

\textit{Real-World Examples Illustrating the Risks}: The presence of ambiguous specifications within the security controls of AI systems elevates the risk of system vulnerabilities. This issue is analogously reflected in the Cloudbleed incident of 2017, wherein a malfunctioning HTML parser chain within Cloudflare's infrastructure led to the unintended exposure of sensitive data \cite{2017MajorTechCrunch}. This particular incident, akin to potential vulnerabilities within AI systems, was magnified by the lack of clear descriptions regarding system behaviors and inadequate testing protocols. It underscores the necessity for precise, well-articulated technical specifications and robust testing protocols to mitigate the threat of security vulnerabilities within AI technologies.

\subsubsection{\textbf{Data vulnerability}}
Our analysis has identified \emph{16} major security concerns, with a focus on the management of sensitive data. The primary issue at hand is the lack of secure and ethical guidelines for managing sensitive data. Without clear data management protocols, AI systems are at substantial risk of privacy violations, unauthorized data access, and misuse, especially concerning data disaggregated on sensitive attributes. This absence of protocols does not merely represent a technical oversight but a profound legal and ethical challenge. It underscores the urgent need for comprehensive data governance frameworks. Such frameworks must ensure that sensitive data is handled with the utmost security and ethical consideration, adhering strictly to privacy laws.

\textit{Real-World Examples Illustrating the Risks}: The 2015 Anthem Data Breach serves as a cautionary tale for AI systems that manage large-scale personal data. In this incident, hackers exploited substandard data protection measures to access sensitive information, illustrating the need for robust security in AI systems \cite{2014EBayPost}. This breach highlights the importance of integrating advanced data encryption and implementing rigorous security protocols in AI systems. Such measures are essential to safeguard against similar vulnerabilities, protecting sensitive data that AI systems often process and store. Additionally, the SolarWinds breach, resulting from vulnerabilities in third-party integrations, points out the risks involved in incorporating external components into AI systems. This incident underlines the importance of vetting third-party AI components and data sources, ensuring they meet stringent security standards to prevent potential breaches. 

\subsubsection{\textbf{Unenforceable security control}} 
We identified \emph{10} security concerns relating to the difficulty in enforcing specific security measures or controls, which can lead to overlooked vulnerabilities.  For example, the lack of clear guidance on documenting and reviewing the use and effectiveness of transparency tools can lead to subpar transparency, potentially causing overlooked security concerns. Similarly, without precise guidance on establishing relevant policies, there could be an improper separation of duties. Additionally, the lack of specificity for ``regular tracking" frequency and method can lead to inconsistent practices, potentially resulting in overlooked issues in human-AI interaction. 

\textit{Real-World Examples Illustrating the Risks}: In 2020, Clearview AI, a facial recognition company, experienced a data breach where its entire client list was stolen \cite{OfficeofthePrivacyCommissionerofCanada2021PIPEDACanada}. This incident highlighted security control issues stemming from unenforceable and inadeqaute data protection practices.  The breach exposed not only the company's client data but also raised concerns about the security of the billions of facial images Clearview AI had scraped from the internet. Without enforceable policies and clear guidelines on how to secure sensitive data, the company left its system vulnerable to unauthorized access, resulting in data exposure and privacy concerns.

\subsection{Expert recommendations}
\textbf{E1} raises awareness of the dangers of under-defined processes in AI systems and advocates for the creation of explicit operational protocols, particularly those that address ethical and social implications, to prevent situations like the controversy over Facebook's News Feed algorithm. E1 calls for stringent quality assurance and risk assessment protocols in high-stakes AI applications, such as autonomous driving technology, referencing the Uber self-driving car incident as a learning opportunity.
\textbf{E2} emphasizes the need to combat ambiguous specifications within AI systems, pointing to the Cloudbleed incident as an example of the risks posed by unclear definitions. E2 recommends the standardization of terminologies and the implementation of comprehensive testing protocols, including boundary and stress testing, to ensure system resilience against unexpected operational scenarios.

\textbf{E3}, focusing on data vulnerabilities, proposes the adoption of a zero-trust architecture, robust data encryption, and stringent access controls. E3 suggests these are key steps for systems that handle sensitive data, drawing parallels to the Anthem data breach to underscore the necessity of such measures. E3 also advocates for continuous monitoring and anomaly detection to address data breaches or unauthorized access attempts swiftly.
\textbf{E4} points out the need for enforceable security measures and controls, taking lessons from the Uber incident and the broader challenge of  ``automation complacency". E4 suggests that clear documentation, regular review mechanisms for transparency tools, and specific guidance on security control enforcement can reduce overlooked vulnerabilities and enhance overall system security.

\section{Quantitative Analysis of Security Standards}
\label{sec:quantitative_analysis}
\begin{figure}[htbp]
\centering
\includegraphics[width=0.99\linewidth]{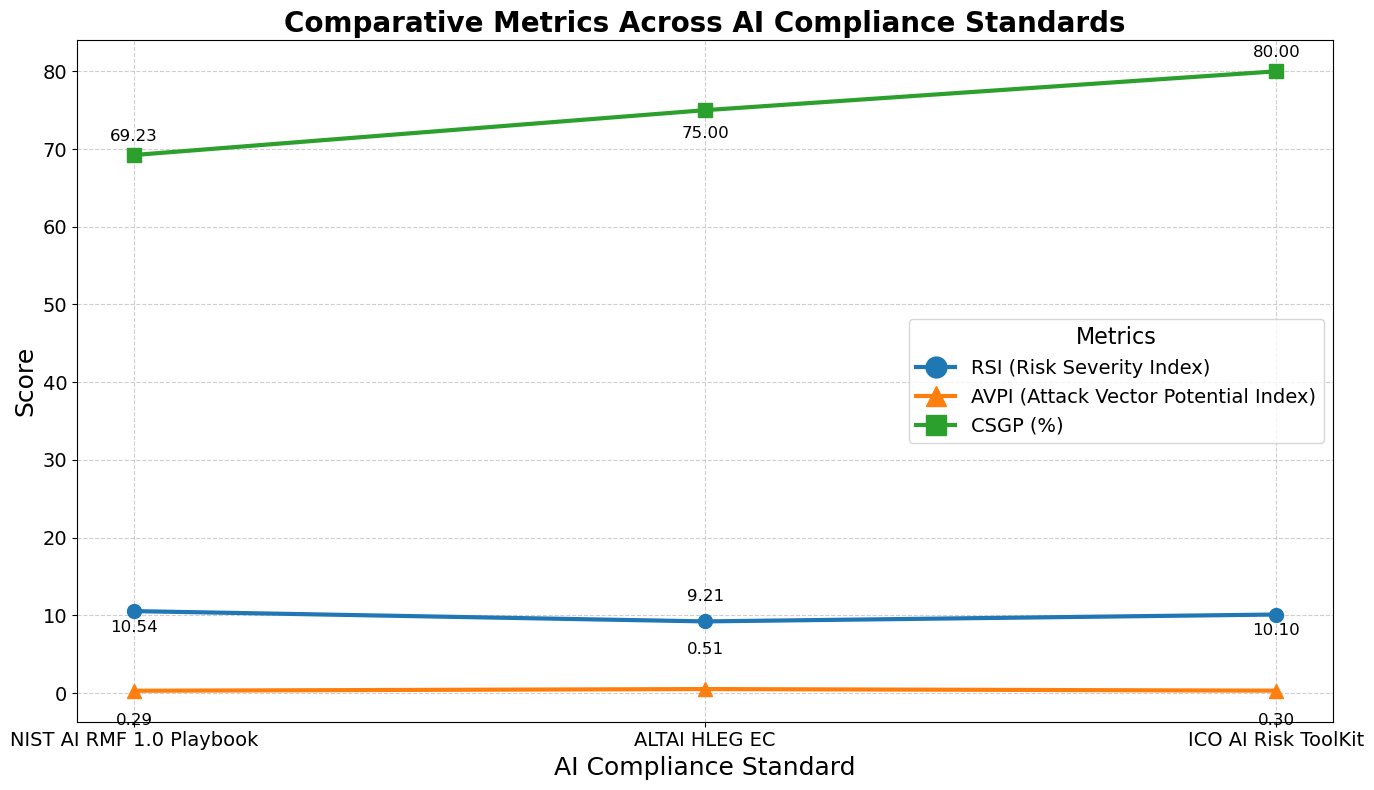}
\caption{\footnotesize Comparative Metrics Across AI Compliance Standards.}
\label{fig:metrics_comparison}
\end{figure}

Our quantitative analysis assesses the effectiveness of AI compliance standards in mitigating security vulnerabilities through five key metrics: Risk Severity Index (RSI), Attack Vector Potential Index (AVPI), Compliance-Security Gap Percentage (CSGP), Total Concerns, and Root Cause Vulnerability Score (RCVS). Table \ref{tab:key_metrics} provides a comparative summary of these metrics across the \textit{NIST AI RMF 1.0 Playbook}, \textit{ALTAI HLEG EC}, and the \textit{ICO AI Risk Toolkit}, while Figure \ref{fig:metrics_comparison} illustrates their relative performance. The analysis highlights that adherence to these frameworks alone does not guarantee security, as significant gaps persist, leaving AI systems vulnerable to various threats.

\begin{table}[h!]
\centering
\caption{Comparative Metrics for RSI, AVPI, CSGP, Total Concerns, and Average RCVS}
\label{tab:key_metrics}
\resizebox{\columnwidth}{!}{%
\begin{tabular}{|l|c|c|c|c|c|}
\hline
\textbf{Standard} & \textbf{RSI} & \textbf{AVPI} & \textbf{CSGP (\%)} & \textbf{Total Concerns} & \textbf{RCVS} \\ \hline
NIST AI RMF 1.0 Playbook & 10.54 & 0.29 & 69.23 & 78 & 0.25 \\ \hline
ALTAI HLEG EC & 9.21 & 0.51 & 75.00 & 28 & 0.33 \\ \hline
ICO AI Risk Toolkit & 10.10 & 0.30 & 80.00 & 30 & 0.25 \\ \hline
\end{tabular}%
}
\end{table}

\subsection{Analysis of Key Metrics}

The Risk Severity Index (RSI) values, ranging from 9.21 to 10.54, indicate moderate-to-high risk severity across all three frameworks. The \textit{NIST AI RMF 1.0 Playbook} reports the highest RSI (10.54), reflecting a greater concentration of severe risks, while the \textit{ALTAI HLEG EC} records the lowest RSI (9.21). This lower RSI for ALTAI reflects its emphasis on high-level principles over detailed, enforceable guidance. However, a lower RSI does not equate to better security, as ALTAI still demonstrates elevated risks across other metrics. NIST’s higher Total Concerns (78) aligns with its broader coverage, while ALTAI's lower Total Concerns (28) reflects its principle-driven, narrower scope.

The Attack Vector Potential Index (AVPI) captures each framework's exposure to attack vectors. The \textit{ALTAI HLEG EC} exhibits the highest AVPI (0.51), suggesting that its reliance on abstract principles leaves exploitable gaps in process definitions. By comparison, \textit{NIST} (0.29) and the \textit{ICO AI Risk Toolkit} (0.30) show slightly lower AVPI values, indicating marginally better mitigation of attack vectors. However, these figures remain concerning, as even the "better-performing" frameworks fail to fully address key vulnerabilities. These AVPI scores highlight that none of the frameworks offers comprehensive protection against potential attack vectors, despite their growing adoption as security guidance standards.

The Compliance-Security Gap Percentage (CSGP) metric reveals critical deficiencies across all three frameworks. The \textit{ICO AI Risk Toolkit} records the highest CSGP (80.00\%), indicating that 80\% of high-risk issues remain unaddressed. The \textit{ALTAI HLEG EC} follows with a CSGP of 75.00\%, while the \textit{NIST AI RMF 1.0 Playbook} reports the lowest CSGP (69.23\%). Although NIST’s relatively lower CSGP reflects a smaller proportion of unresolved concerns, it still points to significant gaps in addressing high-risk issues. ICO's high CSGP suggests that its primary focus on privacy and compliance has not translated into effective security measures. These findings underscore a systemic limitation in compliance-focused frameworks: they provide broad, general guidance but lack the clear, enforceable requirements needed to ensure robust security.

\subsection{Root Cause Analysis}

The Root Cause Vulnerability Score (RCVS) highlights the concentration of vulnerabilities and identifies the primary causes driving unresolved security issues. Among the three frameworks, the \textit{ALTAI HLEG EC} records the highest RCVS (0.33), indicating that a significant proportion of its vulnerabilities stem from specific areas, particularly under-defined processes. This result reflects ALTAI’s reliance on broad principles rather than specific, enforceable guidelines. The elevated RCVS for ALTAI aligns with its high AVPI (0.51), suggesting that these process-related vulnerabilities are not merely theoretical but represent concrete exploitation paths.

In comparison, the \textit{NIST AI RMF 1.0 Playbook} and the \textit{ICO AI Risk Toolkit} both exhibit lower RCVS scores (0.25), reflecting a more distributed set of vulnerabilities. While NIST addresses a broader range of root causes, critical security issues remain unresolved, as evidenced by its CSGP of 69.23\%. Similarly, ICO’s guidance, though extensive, leaves 80\% of high-risk concerns unaddressed, particularly in areas such as data protection, third-party risks, and implementation guidance. These findings underscore a persistent challenge: frameworks designed for general guidance often lack the specificity required to implement critical security controls effectively.

\begin{figure}[htbp]
\centering
\includegraphics[width=\columnwidth]{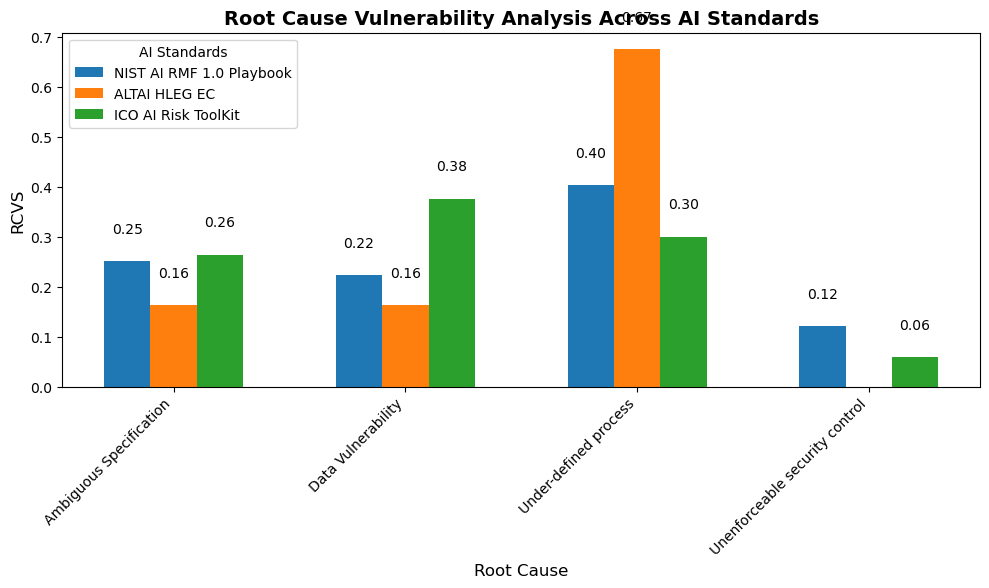}
\caption{\footnotesize Root Cause Vulnerability Analysis Across AI Standards.}
\label{fig:root_cause_analysis}
\end{figure}

The root cause analysis directly addresses the research question: \textit{How effectively do current AI compliance standards address security vulnerabilities when adopted as security guidance frameworks?} The findings reveal that none of the frameworks provides comprehensive protection. The \textit{ALTAI HLEG EC} framework’s reliance on broad principles amplifies risks associated with under-defined processes. In contrast, the \textit{NIST AI RMF 1.0 Playbook} and the \textit{ICO AI Risk Toolkit} offer broader coverage but fail to implement enforceable controls for critical areas such as third-party risks and implementation guidance. These results highlight a fundamental limitation: existing frameworks, regardless of their scope, lack the specificity necessary to mitigate critical security vulnerabilities effectively.

\subsection{Key Insights and Implications}

This analysis uncovers significant shortcomings in AI compliance standards. The \textit{Risk Severity Index} (RSI) and Total Concerns metrics indicate that frameworks like the \textit{NIST AI RMF 1.0 Playbook} provide broader coverage but still fail to address high-risk concerns, as evidenced by its 69.23\% Compliance-Security Gap Percentage (CSGP). In contrast, the principle-driven approach of the \textit{ALTAI HLEG EC} results in fewer overall concerns (28) but concentrates unresolved risks in specific areas, such as under-defined processes, as reflected by its high RCVS (0.33) and AVPI (0.51). The \textit{ICO AI Risk Toolkit} demonstrates an even higher CSGP (80.00\%), revealing that a substantial portion of risks remains unaddressed despite its emphasis on compliance and privacy.

The convergence of these metrics underscores a fundamental issue: existing AI compliance frameworks prioritize broad compliance and guidance over enforceable, actionable security measures. This approach risks giving organizations a false sense of security, as unresolved vulnerabilities—evidenced by high CSGP and RCVS scores—leave them exposed to attack vectors and operational failures. To overcome these challenges, frameworks must move beyond generalized principles and adopt prescriptive security controls designed to directly mitigate high-risk vulnerabilities.

\section{Discussion and Recommendations}
\label{sec:discussion_recommendations}
Our audit of the NIST AI RMF 1.0, ALTAI HLEG EC, and ICO AI Risk Toolkit standards revealed significant security gaps that could expose organizations to vulnerabilities if implemented without modification. Using our novel metrics—Risk Severity Index (RSI), Attack Vector Potential Index (AVPI), Root Cause Vulnerability Score (RCVS) and Compliance-Security Gap Percentage (CSGP)—we quantitatively demonstrated that compliance alone does not guarantee security.
\subsection{Key Vulnerabilities and Recommendations}
The analysis revealed several key vulnerabilities across the audited standards. Notably, under-defined processes emerged as a primary concern, with ALTAI HLEG EC recording a high Root Cause Vulnerability Score (RCVS) of 8.50, and NIST AI RMF 1.0 following with an RCVS of 5.79. These high scores indicate significant gaps in implementation clarity, potentially leading to inconsistent application of security measures. To address this, we recommend introducing specific procedural guidelines that bridge the gap between identifying vulnerabilities and implementing effective controls.
Furthermore, NIST AI RMF 1.0 exhibited a substantial compliance-security gap, with a CSGP of 56.41\%. This alarming figure indicates that over half of the identified risks remain unaddressed despite compliance, underscoring the critical need for mandatory security controls that specifically target AI-related risks such as data poisoning and adversarial attacks.
Data vulnerabilities also emerged as a significant concern, with 31 cases identified across all standards. This finding emphasizes the urgent need for stronger data protection guidelines within AI compliance frameworks. Additionally, ALTAI HLEG EC demonstrated the highest vulnerability to attack vectors, with an AVPI of 4.23. To mitigate this risk, we recommend incorporating comprehensive adversarial threat modeling and implementing robust safeguards against external inputs.

\subsection{Actionable Implications for Real-World AI Systems}
The integration of RSI, AVPI, RCVS and CSGP into organizational security frameworks offers a powerful approach for conducting targeted risk assessments. These metrics enable the identification of vulnerabilities that may not be apparent through standard compliance checks. Policymakers can leverage these metrics to prioritize revisions to existing standards, ensuring that compliance frameworks address the most critical AI security challenges. By embedding these metrics into compliance audits, organizations can transition from reactive responses to proactive strategies that mitigate risks before they materialize.

\section{Conclusion}
\label{sec:conclusions}
Our analysis of three major AI compliance frameworks—NIST AI RMF 1.0, ICO AI Risk Toolkit, 
and ALTAI—reveals that compliance does not necessarily equate to security. We identified 
136 security concerns tied to data vulnerabilities, ambiguous specifications, and unenforceable controls, 
highlighting systemic flaws in how these frameworks address adversarial threats. 
By introducing novel metrics (RSI, AVPI, CSGP, RCVS), we have quantified the severity and root causes 
of these gaps, offering a basis for comparing and improving AI standards.
To support our findings, we have compiled thematic tables of security concern trends and problematic statements in Appendix \ref{app:appendixA}, providing further clarity on the risks identified. Additionally, we have outlined targeted recommendations for the standards in Appendix \ref{sec:appendix_recommendations}. These recommendations are intended to strengthen AI system security by addressing the vulnerabilities uncovered in this study.

\subsection{Future Research Directions}
Future work should focus on expanding the application of our metrics (RSI, AVPI, RCVS and CSGP) to a broader range of international AI governance frameworks. This expansion should involve a larger, more diverse group of experts to ensure the metrics' relevance across various contexts. Developing automated tools for continuous auditing of AI compliance standards is crucial, as it would enable real-time monitoring and allow organizations to assess their security posture as standards evolve.
Further research is needed to explore the delicate balance between specificity and flexibility in compliance standards, ensuring they remain both adaptable and robust. This could involve developing industry-specific guidelines that provide tailored solutions to the unique security challenges faced by different sectors. Longitudinal studies tracking the evolution of AI compliance standards in response to new threats would offer valuable insights into the dynamics of AI governance.
Investigating the effectiveness of our recommendations through case studies and pilot implementations could provide practical guidance for improving AI security governance. Additionally, exploring the integration of these metrics into existing risk management frameworks would enhance their applicability and adoption.
By pursuing these research directions, policymakers and organizations can leverage our approach to enhance the security of AI systems, mitigating potential attack vectors and addressing the gaps identified in current standards. This ongoing work will be crucial in ensuring that AI compliance frameworks keep pace with the rapidly evolving landscape of AI technologies and associated security challenges.

\section*{Acknowledgment}
We want to thank our researchers, who diligently reviewed the standards in great detail, and the experts who confirmed our findings and offered invaluable insights and recommendations.

\bibliographystyle{ACM-Reference-Format}
\bibliography{references}

\appendix

\appendix

\section{AI Compliance Standards Selection Process Detail}
\label{app:selection_standards}

This section provides a detailed overview of the AI compliance standards selected for the audit, including their functions, intended audiences, and enforcement mechanisms.

\subsection{NIST AI RMF 1.0}
\textbf{Released by:} National Institute of Standards and Technology (NIST) \\
\textbf{Release Date:} January 26, 2023 \\
\textbf{Purpose:} To offer a comprehensive strategy for globally managing risks associated with AI systems.

\paragraph{Main Functions:}
\begin{itemize}
    \item \textbf{Govern:} Establish governance and accountability for AI risks.
    \item \textbf{Map:} Identify and categorize risks within AI systems.
    \item \textbf{Measure:} Assess risk severity and likelihood.
    \item \textbf{Manage:} Implement risk mitigation or acceptance strategies.
\end{itemize}
\textbf{Relevance and Non-Compliance:} Acts as a global benchmark for AI risk management, emphasizing the potential for increased risk exposure and consequent financial and reputational harm in cases of non-compliance.

\subsection{UK's AI and Data Protection Risk Toolkit}
\textbf{Released by:} Information Commissioner’s Office (ICO) \\
\textbf{Release Date:} 2020 \\
\textbf{Objective:} To assist organizations in mitigating risks from their AI systems, with a particular focus on data protection.

\paragraph{Components:}
\begin{itemize}
    \item Auditing tools and procedures for risk assessment.
    \item Guidance on AI and data protection laws.
    \item Support resources for ensuring compliance.
\end{itemize}
\textbf{Target Audience and Enforcement:} Aimed at data protection officers and IT professionals, featuring a rigorous enforcement regime with penalties, such as a £7.5 million fine imposed on Clearview AI for non-compliance.

\subsection{European Union's ALTAI}
\textbf{Released by:} European Union \\
\textbf{Purpose:} To provide guidelines for the ethical use of AI technologies.

\paragraph{Key Provisions:}
\begin{itemize}
    \item Principles for human oversight and technical safety.
    \item Guidelines on privacy, data governance, and transparency.
    \item Measures to ensure diversity, non-discrimination, and societal well-being.
    \item Accountability mechanisms for ethical AI use.
\end{itemize}
\textbf{Application and Non-Compliance:} Covers a broad spectrum of entities within the EU, endorsing ethical AI practices with strict measures for non-compliance, including fines and AI use restrictions.

\section{Detailed Participant Information}
\label{app:participant_details}

This study assembled a team of researchers and experts, aligning with best practices in empirical research \cite{Potts1993Software-EngineeringRevisited}. We employed purposive sampling to recruit five researchers and four Subject Matter Experts (SMEs) from diverse backgrounds in academia, industry, and government.

\textbf{Researcher Recruitment:} Researchers were selected based on their proficiency in compliance protocols, involvement in AI system development, and relevant professional experience. They were recruited through LinkedIn and professional networks, excluding the authors of this paper.

\textbf{Expert Recruitment:} SMEs were recruited via professional networks and snowball sampling, aiming to enhance AI security and compliance standards \cite{Largent2017PayingForward, Stevens2022AboveMandates}. This approach allowed us to access a broader network of specialists who provided critical evaluations of our findings.

\textbf{Participant Demographics:} 

The detailed demographics of our participants, including their roles, years of experience, and employment sectors, are presented in Table \ref{tab:detailed_demo}, showcasing the diverse expertise brought to this study by both researchers and experts.

\begin{table}[!ht]
\renewcommand{\arraystretch}{1.3}
\caption{Detailed Demographics of Researchers and Experts}
\label{tab:detailed_demo}
\centering
\begin{tabular}{|c|c|c|c|}
\hline
\rowcolor{gray!20} \textbf{Participant\textsuperscript{1}} & \textbf{Field/Role\textsuperscript{2}} & \textbf{Experience (yrs)} & \textbf{Employment\textsuperscript{3}} \\
\hline
R1 & --- & 27 & G \\
R2 & --- & 8 & A \\
R3 & --- & 10 & A, I \\
R4 & --- & 27 & A \\
R5 & --- & 15 & I \\
\hline
E1 & IT, C & 27 & I, G \\
E2 & IT, SA/PM, C & 36 & I, G \\
E3 & AR, C, ML & 10 & A \\
E4 & IT, ML, RM & 28 & I, G \\
\hline
\end{tabular}
\begin{flushleft}
\footnotesize
\textsuperscript{1} \textit{R1-R5: Researchers, E1-E4: Experts} \\
\textsuperscript{2}  \textit{IT: Information Technology, C: Cybersecurity, PM: Project Manager, AR: Academic Researcher, ML: Machine Learning, RM: Risk Management }\\
\textsuperscript{3} \textit{A: Academia, G: Government, I: Industry}
\end{flushleft}
\end{table}

\textbf{Ethical Considerations:} This study received ethical approval from the University Research Ethics Board. All participants provided informed consent. In line with similar studies \cite{Stevens2022AboveMandates}, experts were not compensated for their participation, which aligns with research suggesting that compensation does not significantly affect participants' responses in certain contexts \cite{Largent2017PayingForward}.

\section{Expert Validation Survey}
\label{app:consentform}

\subsection{Consent Form}
\label{app:c_form}
The participant is presented with the following consent form. Please check all that apply (you may choose any number of these statements):

\begin{itemize}
  \item \textbf{I confirm that I am 18 years or older.}
  \item \textbf{I confirm that I have read and understood this consent form.}
  \item \textbf{I voluntarily agree to participate in this research and want to continue with the survey.}
\end{itemize}

\subsection{Survey Overview}
\label{app:surveyquestions}
This survey asks you to assess the validity of an independent evaluation of \textbf{[standard name]} for the selected subset of questions. An independent evaluation refers to an objective assessment conducted by external experts. Please be as candid and detailed as possible in your responses.

\subsection{Survey Questions}

Please provide your input on the following aspects for each security concern identified:

\begin{enumerate}
  \item \textbf{Organizational Vulnerability:} If your organization strictly adheres to the standard without additional measures, would it be vulnerable to this issue? (Options: agree/plausible/no)
  
  \item \textbf{Likelihood of Exploitation:} If the answer to the first question is "yes" or "plausible," what is the likelihood of this vulnerability being exploited if the standard is followed as written? (Options: Frequent - often occurs, continuously experienced; Likely - occurs several times; Occasional - occurs sporadically; Seldom - unlikely, but could occur at some time; Unlikely - can assume it will not occur)
  
  \item \textbf{Severity of Exploitation:} What would be the severity of exploitation if the standard is followed as written? (Options: Catastrophic - complete system loss, major property damage, full data breach, corruption of all data; Critical - major system damage, property damage, data breach, corruption of sensitive data; Moderate - minor system damage, minor property damage, partial data breach; Negligible - minor system impairment)
  
  \item \textbf{Recommendations:} Based on your experience, what are your recommendations for addressing these security concerns?
  
  \item \textbf{Additional Mitigations:} What additional policies, procedures, or defensive techniques does your organization employ to mitigate this issue?
\end{enumerate}

Please provide your responses to each of these questions for each standard.

\section{Audit Findings}
\label{app:auditfindings}

All our audit findings can be accessed at the following link: \url{https://bit.ly/3L854PI}. The spreadsheet includes three tabs:

\begin{enumerate}
  \item Tab 1: NIST AI RMF 1.0 concerns
  \item Tab 2: ALTAI concerns
  \item Tab 3: AI and Data Protection Risk Toolkit
\end{enumerate}

\section{Visualizations}
\label{app:visualizations}

This section presents additional visualizations that contribute to our research findings. These visualizations provide valuable insights and enhance our understanding of the security concerns of AI compliance standards.

\subsection{Security Concern Impact Levels Matrix}
\label{app:impactmatrix}

Table \ref{tab:risk_matrix} depicts the security concern impact levels derived from the risk mitigation process based on the CRM framework from the U.S. Army risk assessment \cite{Vanvactor2007RiskAssessment}. Levels were assigned according to a CRM risk-assessment matrix, incorporating both probability of occurrence and impact severity levels.

\begin{table}[htbp]
\centering
\caption{Risk Assessment Matrix}
\label{tab:risk_matrix}
\resizebox{\columnwidth}{!}{%
\begin{tabular}{|c|c|c|c|c|c|c|}
\hline
\multicolumn{2}{|c|}{} & \multicolumn{5}{c|}{\textbf{Probability}} \\
\hline
\multicolumn{2}{|c|}{} & Frequent A & Likely B & Occasional C & Seldom D & Unlikely E \\
\hline
\multirow{5}{*}{\textbf{Severity}} & \textbf{Catastrophic I} & E & E & H & H & M \\
\cline{2-7}
& \textbf{Critical II} & E & H & H & M & L \\
\cline{2-7}
& \textbf{Marginal III} & H & M & M & L & L \\
\cline{2-7}
& \textbf{Negligible IV} & M & L & L & L & L \\
\hline
\multicolumn{7}{|l|}{
    \begin{tabular}[c]{@{}l@{}}
    E -- Extremely High Risk \\
    H -- High Risk \\
    M -- Moderate Risk \\
    L -- Low Risk \\
    \end{tabular}
} \\
\hline
\end{tabular}%
}
\end{table}

\subsection{Root Cause Distribution Figures}
\label{app:root_cause_figs}

\begin{figure}[htbp]
    \centering
    \includegraphics[width=0.5\textwidth]{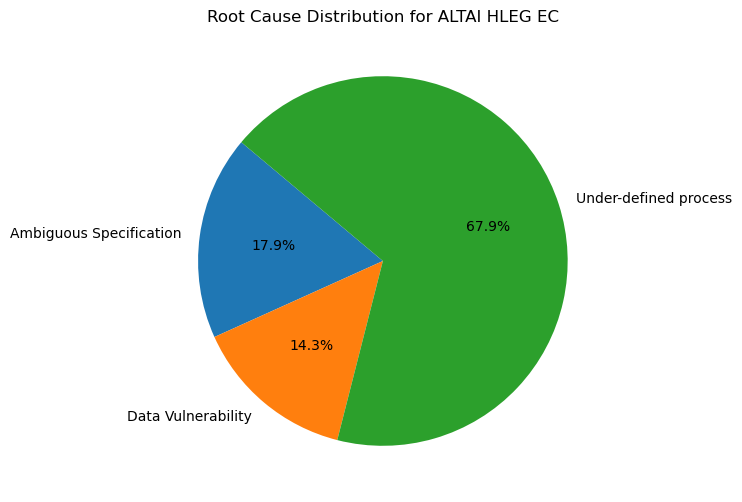}
    \caption{Root Cause Distribution for ALTAI HLEG EC}
    \label{fig:altai_root}
\end{figure}

\begin{figure}[htbp]
    \centering
    \includegraphics[width=0.5\textwidth]{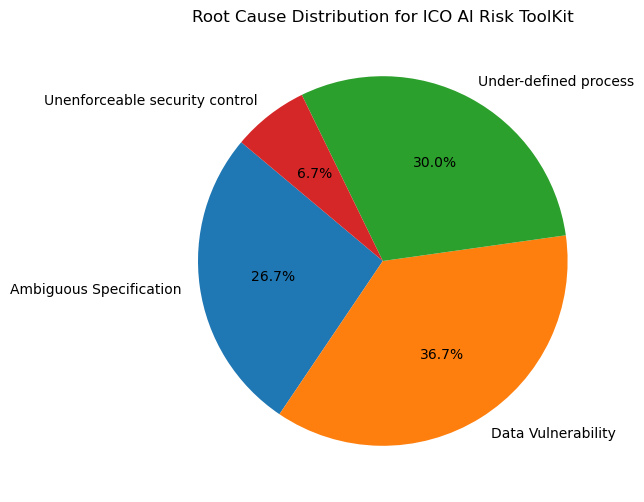}
    \caption{Root Cause Distribution for ICO AI Risk Toolkit}
    \label{fig:ico_root}
\end{figure}

\begin{figure}[htbp]
    \centering
    \includegraphics[width=0.5\textwidth]{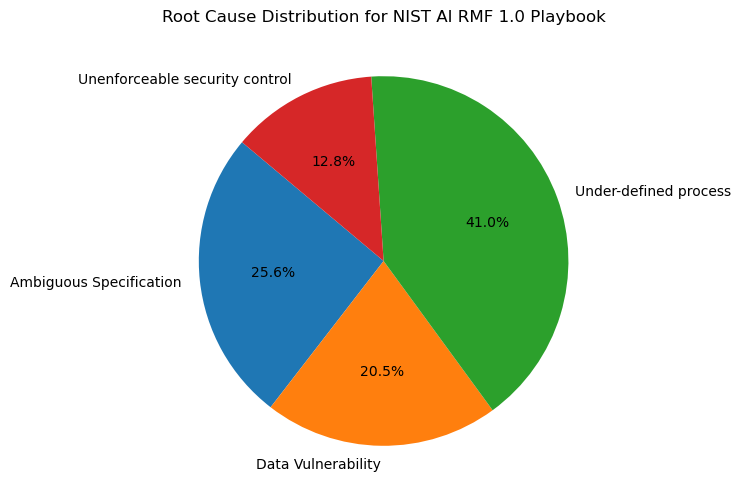}
    \caption{Root Cause Distribution for NIST AI RMF 1.0 Playbook}
    \label{fig:nist_root}
\end{figure}
\subsection{Overall Security Controls Heatmap}
\label{app:heatmaps}

\begin{figure}[ht]
    \centering
    \includegraphics[width=0.5\textwidth]{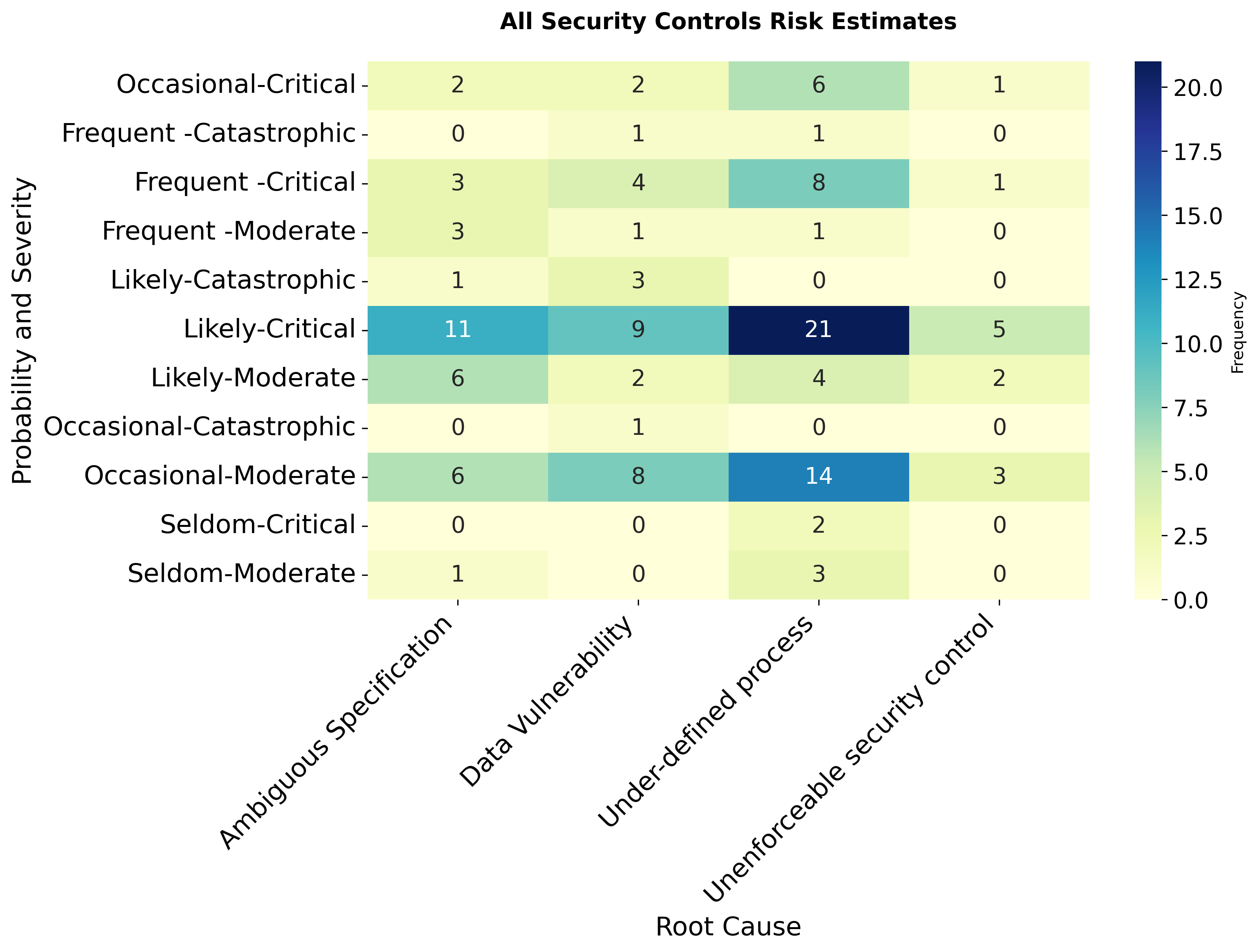}
    \caption{All Standards Security Concerns Heatmap}
    \label{fig:heatmap2}
\end{figure}

Figure \ref{fig:heatmap2} illustrates the overall security controls heatmap, providing an overview of the security landscape across all standards.

\subsection{Risk Matrices}
\label{app:riskmatrices}

This subsection presents individual risk matrices for each audited standard, providing a detailed analysis of risk distribution and impact.

\begin{figure}[ht]
    \centering
    \includegraphics[width=0.5\textwidth]{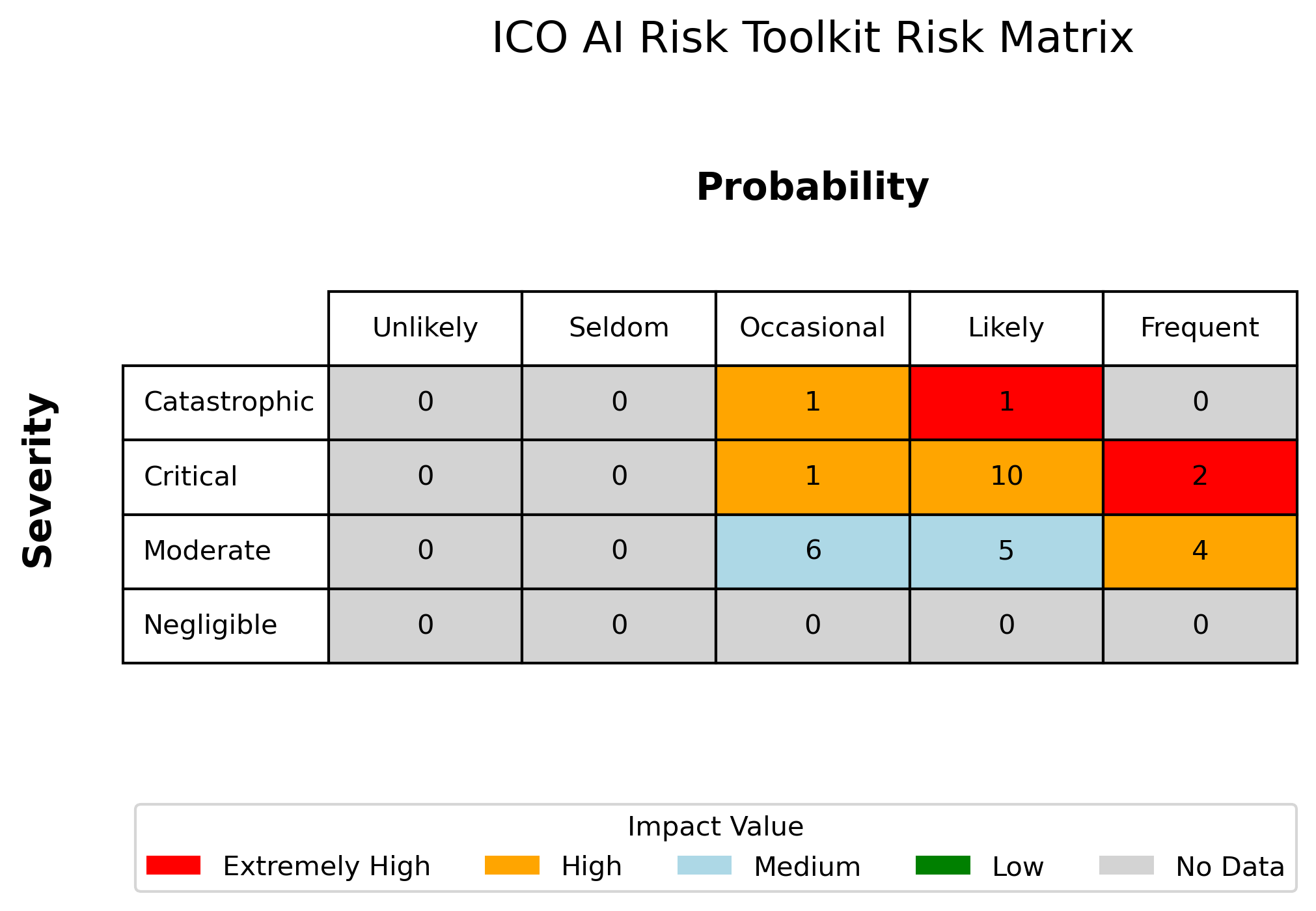}
    \caption{Risk Matrix for ICO AI and Data Protection Toolkit}
    \label{fig:risk_matrix_ico}
\end{figure}

\begin{figure}[ht]
    \centering
    \includegraphics[width=0.5\textwidth]{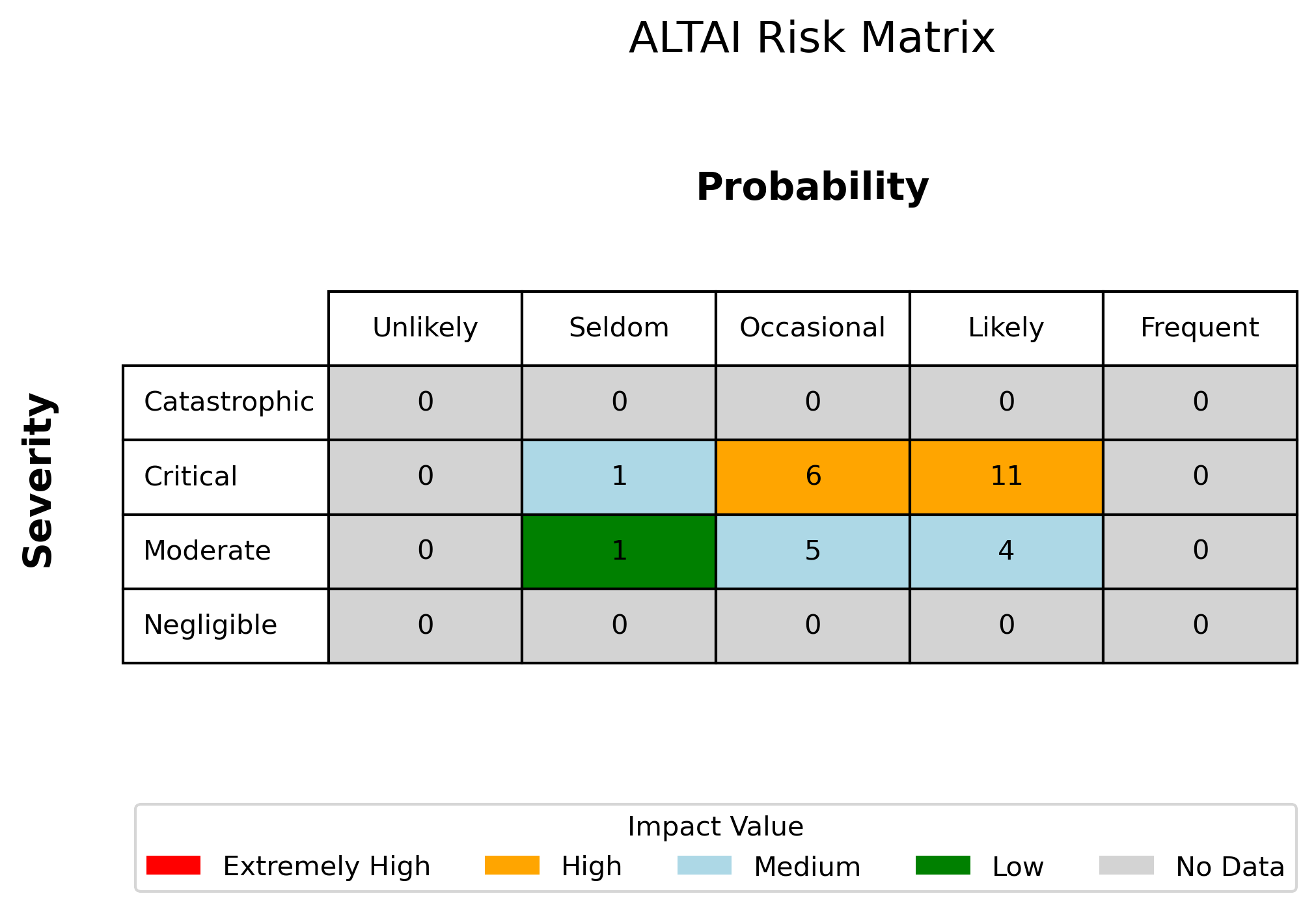}
    \caption{Risk Matrix for Assessment List for Trustworthy Artificial Intelligence (ALTAI)}
    \label{fig:risk_matrix_altai}
\end{figure}

\begin{figure}[ht]
    \centering
    \includegraphics[width=0.5\textwidth]{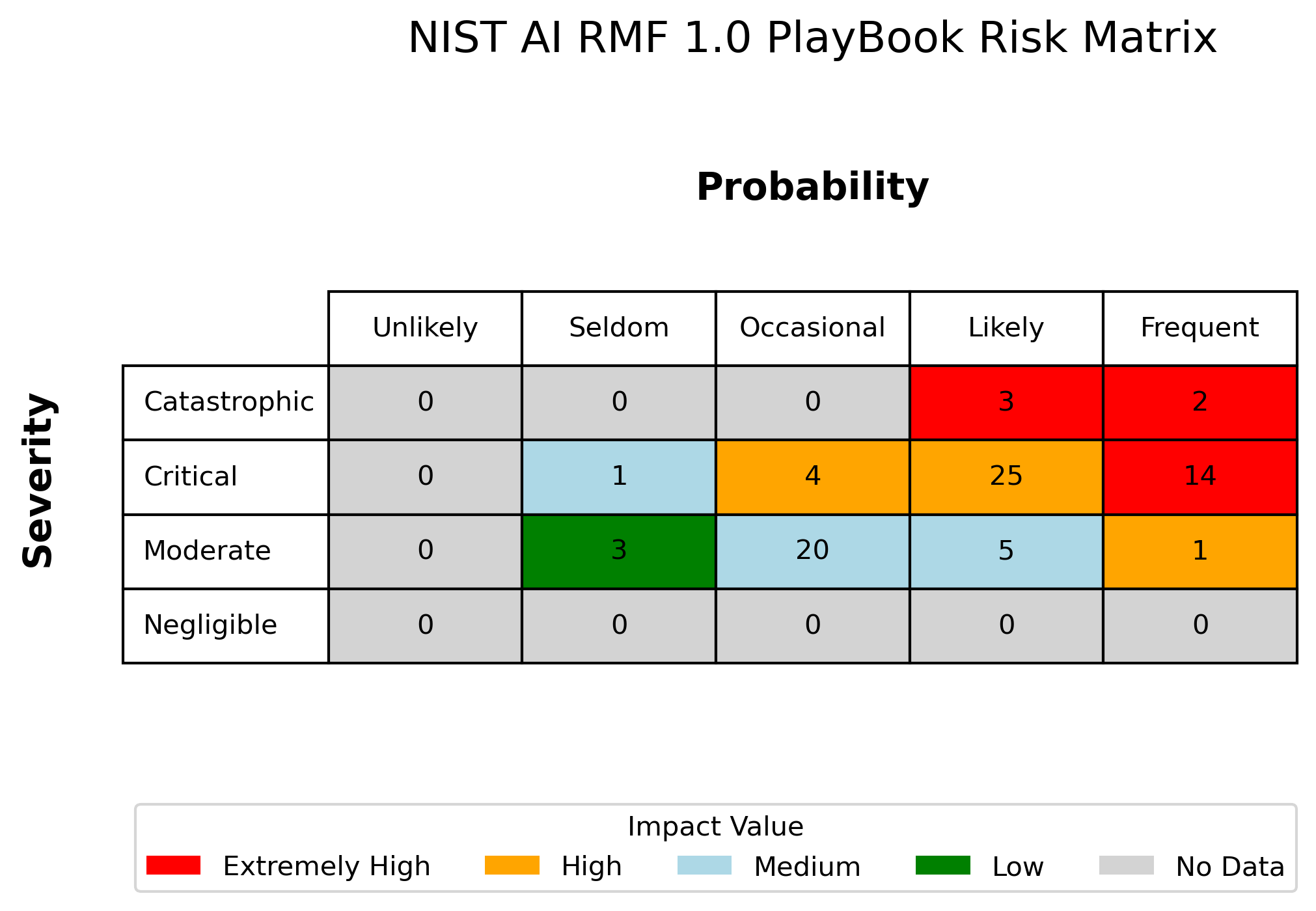}
    \caption{Risk Matrix for NIST Artificial Intelligence Risk Management Framework (AI RMF)}
    \label{fig:risk_matrix_rmf}
\end{figure}

\begin{itemize}
  \item \textbf{Figure \ref{fig:risk_matrix_ico}}: Risk Matrix for ICO AI and Data Protection Toolkit.
  \item \textbf{Figure \ref{fig:risk_matrix_altai}}: Risk Matrix for Assessment List for Trustworthy Artificial Intelligence (ALTAI).
  \item \textbf{Figure \ref{fig:risk_matrix_rmf}}: Risk Matrix for NIST Artificial Intelligence Risk Management Framework (AI RMF).
\end{itemize}

These visualizations provide a comprehensive visual representation of the identified security concerns and their associated risk levels, aiding in the interpretation and analysis of our research findings. The detailed examination of these visualizations supports the assessment of AI compliance standards and offers valuable insights for potential improvements and future research endeavors.

\section{Detailed Analysis of Security Concern Trends}
\label{app:appendixA}

This section presents a comparative analysis of the security concerns identified across three major AI compliance standards: the ICO AI Toolkit, the European Union's Assessment List for Trustworthy Artificial Intelligence (ALTAI), and the NIST AI Risk Management Framework (AI RMF). By examining these standards, we aim to highlight common challenges and issues, facilitating a deeper understanding of their implications for AI security and data protection.

\subsection{Overview of Security Concern Trends}
Our analysis categorizes the security concerns into key themes that represent overarching issues affecting AI system security and compliance. The comparison of these themes across the three standards reveals areas where each standard excels or falls short.

\begin{table*}[htbp]
\centering
\caption{Comparison of Security Concern Themes Across AI Standards}
\label{tab:security_concern_themes}
\small
\begin{tabular}{|p{0.22\textwidth}|p{0.25\textwidth}|p{0.25\textwidth}|p{0.25\textwidth}|}
\hline
\rowcolor{gray!20} \textbf{Security Concern Theme} & \textbf{ICO AI Toolkit} & \textbf{ALTAI (EU)} & \textbf{NIST AI RMF} \\
\hline
\textbf{Data Protection} & 
Lacks detailed guidelines for implementation. Focuses on general principles rather than specifics. & 
Vague on practical application of data protection. Insufficient procedural guidance. & 
Requires more robust and detailed measures. Provides general frameworks without depth. \\
\hline

\textbf{Accountability} & 
Roles and responsibilities are poorly defined. Lacks specificity in accountability mechanisms. & 
Insufficient detail on enforcing accountability. Broadly defined roles with no clear action points. & 
Needs clearer definitions of roles. Requires detailed allocation of responsibilities. \\
\hline

\textbf{Third-Party Risks} & 
Minimal focus on third-party risk assessment. Lacks comprehensive audit guidelines for third parties. & 
No specific guidelines for third-party audits. General mention without actionable steps. & 
Lacks detailed protocols for third-party risk management. Needs clear enforcement strategies. \\
\hline

\textbf{Compliance Requirements} & 
Broad and non-specific recommendations. Lacks actionable compliance steps. & 
General criteria without clear implementation guidelines. Compliance expectations are broadly defined. & 
Contains generic compliance statements. Needs operational specificity for implementation. \\
\hline

\end{tabular}
\end{table*}

Each row in Table \ref{tab:security_concern_themes} represents a specific theme of security concerns, comparing how the ICO AI Toolkit, ALTAI, and NIST AI RMF standards address these issues. This analysis reveals a consistent struggle across all standards to provide detailed, actionable guidelines, particularly in areas critical for operational security like data protection and third-party risk management. The findings underscore the need for these standards to evolve, incorporating more comprehensive protocols and explicit operational steps to ensure that AI systems are secure and compliant across various implementation environments.

\subsection{Analysis of Problematic Statements}
Following the thematic analysis, we identified specific problematic statements within each standard that lack the necessary specificity or clarity to effectively guide AI system security. These statements are categorized by theme, as summarized in Table \ref{tab:problematic_statements}.

\begin{table}[htbp]
\centering
\caption{Summary of Problematic Statements by Standard}
\label{tab:problematic_statements}
\small
\begin{tabular}{|p{0.25\linewidth}|p{0.65\linewidth}|}
\hline
\textbf{Theme} & \textbf{Problematic Statements} \\
\hline
\multicolumn{2}{|c|}{\textbf{ICO AI Toolkit}} \\
\hline
Vagueness & ``Assess risks where necessary" lacks specific guidelines on frequency and methods. \\
\hline
Lack of Specificity & ``Implement security measures" lacks description of required levels or types of measures. \\
\hline
Operational Ambiguities & ``Manage data privacy" lacks clear operational steps or examples. \\
\hline
Enforcement Issues & Provides inadequate guidance on enforcing compliance with privacy laws. \\
\hline
Documentation & ``Document AI processes" without outlining the extent or depth of documentation needed. \\
\hline

\multicolumn{2}{|c|}{\textbf{ALTAI (EU)}} \\
\hline
Vagueness & ``Ensure AI transparency" lacks detailed guidance on implementation. \\
\hline
Lack of Specificity & ``Audit AI systems periodically" without defined intervals or criteria. \\
\hline
Operational Ambiguities & ``Adopt risk management frameworks" lacks detailed application methodologies. \\
\hline
Enforcement Issues & ``Enforce data protection" lacks clear procedures for dealing with violations or breaches. \\
\hline
Documentation & ``Record all data usage decisions" without specifying the required detail. \\
\hline

\multicolumn{2}{|c|}{\textbf{NIST AI RMF}} \\
\hline
Vagueness & ``Use privacy-enhancing technologies" without specifying technologies or strategies. \\
\hline
Lack of Specificity & ``Maintain data integrity" lacks explanation on specific measures. \\
\hline
Operational Ambiguities & ``Ensure system resilience" does not specify standards or benchmarks for resilience. \\
\hline
Enforcement Issues & ``Monitor third-party vendors" lacks clear monitoring techniques or compliance requirements. \\
\hline
Documentation & ``Archive all system updates" without guidelines on methods or duration. \\
\hline
\end{tabular}
\end{table}

The issues highlighted in Table \ref{tab:problematic_statements} emphasize the need for all standards to improve clarity and provide more detailed, actionable instructions. To address these challenges, we recommend focusing on the following areas:

\begin{itemize}
    \item \textbf{Specific Criteria for Risk Assessment and Auditing}: Define clear criteria for risk assessment and auditing to ensure consistency and comprehensiveness.
    \item \textbf{Operational Steps for Data Privacy and Security Measures}: Provide clear, step-by-step guidelines for implementing data privacy and security measures to enhance compliance.
    \item \textbf{Detailed Enforcement and Monitoring Procedures}: Include explicit procedures for enforcement and monitoring, particularly in managing third-party risks.
    \item \textbf{Comprehensive Documentation Requirements}: Establish clear guidelines on the extent and depth of documentation needed to ensure all relevant information is captured and maintained.
\end{itemize}

By addressing these areas, standard-setting bodies can enhance the effectiveness and applicability of compliance frameworks, leading to better-governed and more secure AI systems.

\section{Detailed Recommendations for Standard Improvements}
\label{sec:appendix_recommendations}

\subsubsection{NIST AI RMF 1.0 Playbook}
The NIST AI RMF 1.0 Playbook requires significant improvements to address its high Compliance-Security Gap Percentage (CSGP) and under-defined processes. We recommend enhancing clarity around ambiguous security control guidelines, particularly those related to governance and model retraining. For instance, the standard should provide explicit guidance on documenting and reviewing the use and effectiveness of transparency tools, addressing the vagueness that currently leads to subpar transparency.

Precise guidance on establishing policies for separation of duties is crucial. The standard should also address the lack of specificity in "regular tracking" frequency and methods by providing clear timelines and methodologies for monitoring human-AI interaction. These improvements could help prevent incidents like the Uber self-driving car accident, where unclear processes and insufficient oversight led to a pedestrian fatality \cite{2020UbersNews}. Furthermore, drawing lessons from the Facebook News Feed Algorithm controversy \cite{2022FacebookNewsEngadget}, the standard should mandate stringent quality assurance and risk assessment protocols for high-stakes AI applications.

\subsubsection{ICO AI Risk Toolkit}
The ICO AI Risk Toolkit needs to address data vulnerabilities more comprehensively, especially in areas of data governance and privacy. We recommend mandating, rather than merely suggesting, data flow mapping, addressing the gap identified in Section 1.3 of the standard. This measure could help prevent incidents like the MLFlow vulnerability (CVE-2023-6975), where inadequate data flow oversight led to potential unauthorized access \cite{KevinTownsend2024EightSecurityWeek}.

The toolkit should provide specific data requirements and emphasize data minimization principles, addressing the vagueness found in Section 1.7. This would help prevent incidents like the Meta Pixel controversy, where unclear guidelines led to the unauthorized collection and transmission of sensitive financial information \cite{Fyler2023WhyTechHQ, Hongsdusit2022TaxMarkup}. Additionally, the toolkit should implement more specific guidelines for under-defined processes, such as clearer steps for reporting and managing security breaches, enhancing its overall effectiveness.

\subsubsection{ALTAI HLEG EC}
To mitigate the high Attack Vector Potential Index (AVPI) in the ALTAI HLEG EC standard, we recommend improving definitions and clarity, particularly around ethical AI use and human oversight. The standard should provide clear definitions and implementation guidance for "state-of-the-art" privacy and data protection measures, addressing the vagueness identified in Requirements 2 and 3.

Reducing the attack vector potential requires more prescriptive and scenario-based guidelines. This could help prevent incidents like those demonstrated in Comiter's (2019) research, where AI systems were fooled by inputs crafted to exploit their vulnerabilities \cite{Comiter2019AttackingIt}. Clearer boundaries around human-AI interaction should be established to prevent issues like those seen in the UK Algorithmic Grade Prediction scandal, where lack of transparency led to public outrage over perceived unfair and biased outcomes \cite{2020UKVerge, Collins2021ArtificialAgenda}.

Furthermore, the standard should address the risks associated with over-reliance on AI in critical fields like healthcare. The controversy surrounding IBM's Watson for Oncology in 2018 \cite{2018IBMsOnline, Bernal2022TransparencyWorldwide} highlights the need for transparent decision-making processes and expert validation in AI systems, especially those deployed in sensitive areas.
\end{document}